\documentclass[11pt,english]{article}
\usepackage[utf8]{inputenc}
\usepackage[a4paper]{geometry}
\usepackage{amsmath}
\usepackage{amsthm}
\usepackage{amssymb}
\usepackage{graphicx}
\usepackage{esint}

\makeatletter
\newcommand{\lyxaddress}[1]{
\par {\raggedright #1
\vspace{1.4em}
\noindent\par}
}

\@ifundefined{date}{}{\date{}}
\usepackage[pagebackref = false]{hyperref}
\usepackage{cite}
\usepackage{chemfig}

\makeatother

\usepackage{babel}
\begin{document}

\title{Quantum transport and utilization of free energy in protein $\alpha$-helices}

\author{Danko D. Georgiev\textsuperscript{1} and James F. Glazebrook\textsuperscript{2}}

\maketitle

\lyxaddress{$^{1}$ Institute for Advanced Study, 30 Vasilaki Papadopulu Str.,
Varna 9010, Bulgaria. E-mail: \href{mailto:danko.georgiev@mail.bg}{danko.georgiev@mail.bg}}

\lyxaddress{$^{2}$ Department of Mathematics and Computer Science, Eastern Illinois
University, Charleston, Illinois 61920-3099, USA. E-mail: \href{mailto:jfglazebrook@eiu.edu}{jfglazebrook@eiu.edu}}

\begin{abstract}
The essential biological processes that sustain life are catalyzed by protein nano-engines, which maintain living systems in far-from-equilibrium ordered states. To investigate energetic processes in proteins, we have analyzed the system of generalized Davydov equations that govern the quantum dynamics of multiple amide~I exciton quanta propagating along the hydrogen-bonded peptide groups in $\alpha$-helices. Computational simulations have confirmed the generation of moving Davydov solitons by applied pulses of amide~I energy for protein $\alpha$-helices of varying length. The stability and mobility of these solitons depended on the uniformity of dipole--dipole coupling between amide~I oscillators, and the isotropy of the exciton--phonon interaction. Davydov solitons were also able to quantum tunnel through massive barriers, or to quantum interfere at collision sites. The results presented here support a non-trivial role of quantum effects in biological systems that lies beyond the mechanistic support of covalent bonds as binding agents of macromolecular structures. Quantum tunneling and interference of Davydov solitons provide catalytically active macromolecular protein complexes with a physical mechanism allowing highly efficient transport, delivery, and utilization of free energy, besides the evolutionary mandate of biological order that supports the existence of such genuine quantum phenomena, and may indeed demarcate the quantum boundaries of life.
\end{abstract}

\section{Introduction}

Living organisms are considered as open physical systems which utilize the availability of
free energy to maintain homeostasis, respond to stimuli, adapt to
their environment, grow, reproduce, and to evolve \cite{Oparin1957,Koshland2002,Trifonov2011}. All of these biological functions are implemented by the large-scale participation and interaction of proteins.
The high versatility of protein functions is achieved by linear polymerization
of 20~different standard amino acids into polypeptide chains \cite{Kuhlman2019} (noteworthy, non-standard amino acids also exist, cf.~\cite{Atkins2002}).
The linear sequence of amino acids in polypeptides is often referred to as the primary structure.
The primary structure folds into two main types of hydrogen-bonded secondary structures,
$\alpha$-helices and $\beta$-sheets \cite{Pauling1951a,Pauling1951b}.
Other, special helical secondary structures, such as the $3_{10}$-helix and the
$\pi$-helix, are also found in proteins (for review of their functional importance see \cite{Vieira-Pires2010,Kumar2015}).
The ensuing organization of secondary structures through hydrogen
bonding, ionic bonding, dipole--dipole interactions, London dispersion
forces, or through covalent disulfide bonds, provides each protein with its own tertiary
structure that is uniquely shaped. Further quaternary assembly of multiple
protein subunits provides the means for evolutionary design of nano-engines
equipped with multiple active sites, including an adenosine triphosphate
(ATP) hydrolytic site for release of free energy, an active catalytic
site for converting the released energy into biologically useful work,
and a number of allosteric sites for regulation of protein activity.
This permits utilizing biochemical energy stored in high-energy
ATP pyrophosphate bonds to fuel bio-processes such as protein-assisted
directed motion, synthesis of biomolecules, or for transporting biochemical
substances across lipid membranes.

The primary importance of proteins for life is reflected in their
name: the term ``protein'' comes from the Greek words ``protos'',
meaning ``first'', or ``proteos'', meaning ``first of all''.
In addition to being major catalysts in
biological processes, proteins may indeed have been the constituents of the very first physical systems of life.
While in modern organisms genetic information for
protein production is stored in nucleic acids, recent stochastic simulations
support the protein foldamer hypothesis for abiotic origin of life
\cite{Guseva2017}. In particular, short protein sequences composed
of hydrophobic and polar amino acids were found to collapse into relatively
compact structures with exposed hydrophobic surfaces, which in turn
catalyze the elongation of other such hydrophobic/polar protein polymers
\cite{Guseva2017}. Once the first living replicators were stochastically
assembled, natural selection would have kicked in and selected those
replicators that exploit available energy sources in the most efficient
way for the purposes of reproduction. Eventually, after 3.5 billion years of evolution
\cite{Schopf2018}, modern day organisms are capable of the effective use of energy as
released by single ATP molecules so as to execute highly specialized
biological processes, including single steps of the kinesin motor on cytoskeletal
protein railways \cite{Schnitzer1997,Coy1999}, or phosphorylation of single amino acid residues in voltage-gated ion channels for modulation of the channel electric conductance \cite{Ahn2007,Tucker2002}.

In living systems, energy is transferred in minute quantities because
higher energy densities are detrimental to the delicate and fragile structures
\cite{Szoke2016}. While the behavior of all molecules is fundamentally
described by quantum mechanics, the highly efficient utilization of
energy by proteins suggests that quantum effects may play a non-trivial
role, and one that lies beyond the deterministic, mechanistic structural support of covalent
bonds that bind macromolecules together \cite{Matsuno2006,Matsuno2017,Melkikh2015,Marais2018,Georgiev2017,Georgiev2020,GeorgievGlazebrook2007,GeorgievGlazebrook2012}.
In this present paper we undertake studying the quantum dynamics resulting
from the generalized Davydov Hamiltonian for the transport of energy by
multiple amide~I quanta in protein $\alpha$-helices. We show that
the resulting system of differential equations admits soliton solutions,
for which the corresponding waveforms preserve their shapes during propagation, reflect from protein
ends, tunnel through massive barriers, and interfere quantum mechanically at collision
points to produce sharply focused peaks of concentrated energy. These instances of
quantum phenomena appear to be instrumental in delivering biochemical
energy to active catalytic sites, and moreover, are expected to be indispensable for the theoretical basis
supporting the quantum boundaries of life.
Indeed, in making this contribution to the currently burgeoning science of quantum biology, we are indebted to Schr\"{o}dinger's astonishing insight which predicted quantum mechanical effects as ubiquitous within living systems \cite{Schrodinger1977}. Presently there is evidence of these effects occurring in several cases: photosynthesis (coherence), avian navigation (entanglement), olfaction (tunneling), the kinetic isotope effect in enzymatic reactions (tunneling), as particular instances (reviewed in \cite{Lambert2013,Brookes2017}; see also earlier works such as \cite{Devault1984,Sutcliffe2000,Basran2001,Fleming2011}).

\section{Protein $\alpha$-helix structure and infrared spectra}

Protein polypeptide chains are linear polymers that are assembled
from a repertoire of 20~different standard amino acids joined together through
peptide bonds from N-terminus to C-terminus. The identity of each
amino acid is determined by its side chain, known as an R~group.
The chemical structure of a generic tripeptide is shown below.

\begin{center}
\chemfig{H-[:45]N(-[:90]H)-[:-45]CH(-[:-90]R_1)-[:45]C(=[:90]O)-[:-45]N(-[:-90]H)-[:45]CH(-[:90]R_2)-[:-45]C(=[:-90]O)-[:45]N(-[:90]H)-[:-45]CH(-[:-90]R_3)-[:45]C(=[:90]O)-[:-45]OH}
\end{center}For the construction of protein molecular machines, the flexibility
of their polypeptide chains is essential \cite{Ichinose1991}. This
allows for organizing the polypeptide primary structure into
ordered secondary structural elements such as the protein $\alpha$-helix.
Linus Pauling's prediction of the protein $\alpha$-helix in 1951
is one of the greatest achievements in structural biology \cite{Edison2001}.
For his accomplishments in revealing the three-dimensional geometry
of protein secondary structural elements, Pauling was awarded the
1954 Nobel Prize in Chemistry. Besides revealing the protein mysteries
of life, it is perhaps not surprising that this distinguished fellow was also actively
involved in the preservation of life itself, which subsequently earned him the Nobel Peace Prize in
1962, thus making Pauling the only person to have been awarded two unshared Nobel Prizes \cite{Kovac1999}.

\begin{figure}[t]
\begin{centering}
\includegraphics[width=90mm]{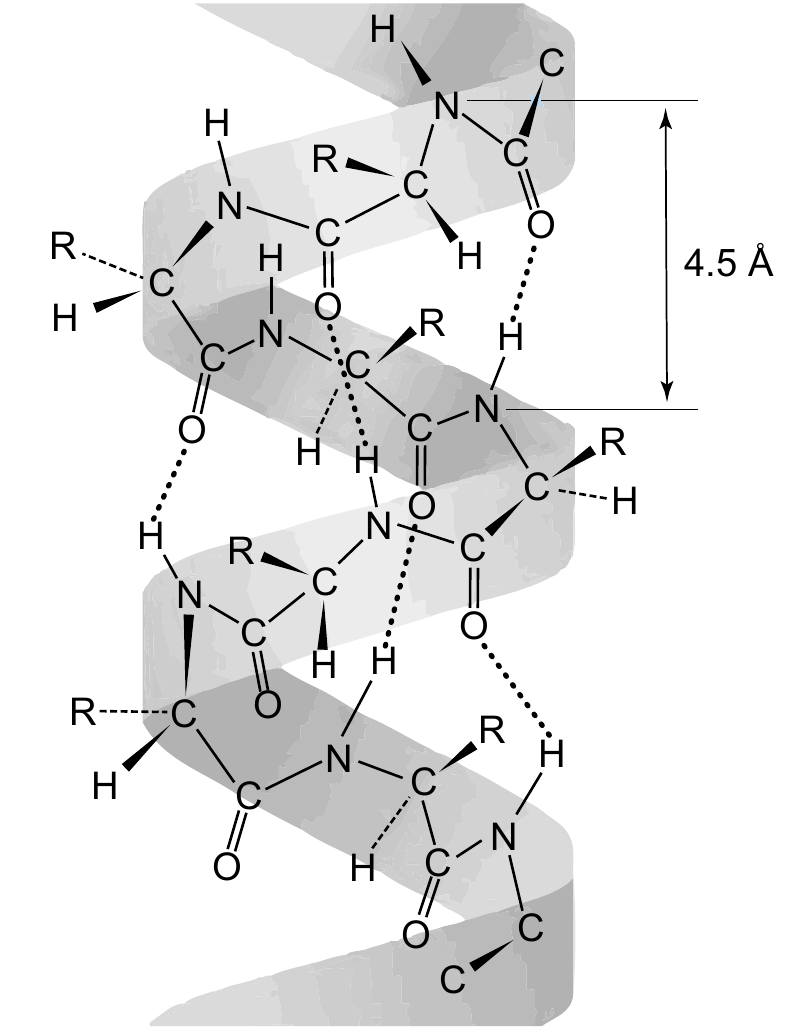}
\par\end{centering}

\caption{\label{fig:1}A protein $\alpha$-helix fragment with 3.6 amino acids
per turn. The helical structure is supported by three chains of hydrogen-bonded
peptide groups $\cdots$H--N--C=O$\cdots$H--N--C=O$\cdots$ referred
to as $\alpha$-helix spines. Modified from Ref.~\cite{GeorgievGlazebrook2019}.}
\end{figure}

Protein $\alpha$-helices are right-handed spirals with 3.6 amino
acids per turn, in which the N--H group of an amino acid is hydrogen-bonded
with the C=O group of the amino acid that appears four residues earlier
in the polypeptide chain (Fig. \ref{fig:1}).
In fact, for $\alpha$-helices there is reliable evidence for the effect of hydrogen bonding on vibrational frequency \cite{Barth2007}.
The helical structure is supported by three chains of hydrogen-bonded peptide groups \mbox{$\cdots$H--N--C=O$\cdots$H--N--C=O$\cdots$} referred to as $\alpha$-helix spines. The typical bond lengths in
the peptide group are as follows: N--H bond length is 101 pm, C--N
bond length is 132 pm, and C=O bond length is 123 pm \cite{Pauling1951a}.
Because of the resonance between C=O and C--N bonds, their bond lengths
are intermediate between a single and a double bond of the corresponding
atoms, and the peptide bonds acquire a planar geometry.
Full quantum simulations of electron molecular orbitals in the protein $\alpha$-helix using Kohn--Sham density functional theory \cite{Sholl2009} are within the capabilities of modern quantum chemistry software applications, but this task would require thousands of hours of running time on a supercomputer \cite{Kolev2013,Kolev2018}. Here, we will mainly rely on experimental data obtained from X-ray crystallography, or from the infrared spectroscopy of proteins.

Generally, hydrogen bonding is regarded as significantly instrumental for stabilizing the secondary structure of proteins in two ways: i) through lowering the frequency of stretching vibrations by reducing restoration forces, and ii) by increasing the frequency of bending through increased restoration \cite{Barth2007,Pace2014}). Under these circumstances, hydrogen bonding can impact the amide functional class by stabilizing its [$^{-}$O--C=N--H$^{+}$] structure over its [O=C--N--H] structure. Significantly, such bonding reliance can be influenced by the form of amide stretch vibrations. Amide~I vibrations mainly arise from the C=O stretching vibrations, while supplemented by minor effects elsewhere \cite{Barth2007}.
As an example, Myshakina \emph{et al}. \cite{Myshakina2008} demonstrated that the frequency shifts of amide~I and amide~III bands (see below) may function as significant regulators for hydrogen bonding at the C=O and N--H sites of certain peptide bonds. Clearly, it is most significant that the proton in such hydrogen bonded systems is a quantum entity. Note that quantum nuclear effects may weaken relatively weak hydrogen bonds, while in contrast they may actually fortify the relatively strong ones \cite{Li2011}. Relevant in this case is the induced fit method for quantum H-bonding, which supports the molecular interactions for inducing conformational transitions in the binding sites of classes of enzymes \cite{Pusuluk2018a}. In such a molecular recognition event, the dynamics of tunneling of electrons of proton-acceptor atoms or protons of hydrogen atoms is also significantly instrumental (see below).
The tunneling of electrons of proton-acceptor atoms or protons of the hydrogen atoms generating quantum correlations have been studied in Refs.~\cite{Pusuluk2018a,Pusuluk2018b}.
Further, in the thermal states of hydrogen bonds, substantial tunneling assisted quantum entanglement can be detected. In particular, if covalent bonding accompanies ionic associates, as is the case for a covalent bond created between electronegative and hydrogen atoms, then quantum
entanglement may be subsequently hypothesized for various instances of ligand binding \cite{Pusuluk2018b}.

Fourier-transform infrared (FTIR) spectroscopy allows for experimental
measurement, and for plotting the absorption of infrared light by sample
material versus the wavelength of the absorbed light. The application
of FTIR spectroscopy to proteins has revealed several absorption bands
that correspond to polypeptide backbone vibrations \cite{Barth2007,Krimm1986,Hamm1998,Manas2000}.

\emph{Amide A}. Among all absorption bands, most energetic is the
amide A band near 3300~cm$^{-1}$ (0.41 eV), which is due to \textgreater{}95\%
N--H stretch. The amount of this energy correspond exactly to the
free energy released from a single ATP molecule. As a result of its
exclusive localization on the N--H group, however, the amide A band
in proteins appears to be insensitive to the secondary structure of
the polypeptide backbone \cite{Barth2007}.

\emph{Amide~I}. Particularly sensitive to the protein secondary structure
is the amide~I band near 1650 cm$^{-1}$ (0.2 eV), which is due to
70-85\% C=O stretch and 10-20\% \mbox{C--N}~stretch \cite{Krimm1986}.
The free energy released from a single ATP molecule is sufficient
to excite two amide~I quanta. Importantly, resonance interaction can
occur between two C=O oscillators when one of them is in an excited
state. For distances over 300 pm, the main contribution to the interaction
energy is due to transition dipole coupling \cite{Krimm1986}. The
fundamental mechanism that renders the amide~I vibration sensitive
to secondary structure is the transition dipole coupling, because
the coupling between the oscillating dipoles of neighboring amide
groups depends upon their relative orientation and their distance
\cite{Barth2007}. The energy absorbed by a given C=O oscillator is
readily transferred to nearby oscillators, which leads to delocalized
excited states \cite{Hamm1998}.

\emph{Amide~II}. Also sensitive to the secondary structure of proteins,
albeit in a less straightforward way, is the amide~II band near 1550
cm$^{-1}$ (0.19 eV), which is due to 40-60\% N--H bend, 18-40\% C--N
stretch, and 10\% C--C stretch.

\emph{Amide~III}. With lowest energy is the amide~III band near 1300
cm$^{-1}$ (0.16 eV), which is due to in-phase combination of 40\%
C--N stretch and 30\% N--H bend.

As an experimental technology, FTIR spectroscopy can be applied to monitor
protein structure in the liquid and dried (lyophilized) state. However,
water can interfere with FTIR measurements of protein samples because
it is strongly absorbent in the amide~I region (how water is distinguished and interpreted in protein chemistry is surveyed in \cite{Wiggins1990,Wiggins2008}).
Consequently, FTIR spectroscopy
is best suited for lyophilized (freeze-dried) protein samples. Measurements
can also be obtained for protein samples in solution, but a high (\textgreater{}3
mg/ml) protein concentration is required \cite{Yang2015}.
We also mention that femtosecond infrared pump-probe spectroscopy has proven to be highly effective for analyzing the amide~I band in relationship to the N--H vibrations \cite{Edler2002b,Edler2002a,Edler2004b,Edler2004a}.

\section{\label{sec:3}The generalized Davydov model of protein $\alpha$-helices}

Based on the strong dependence of amide~I energy on the protein secondary structure,
Alexander Davydov developed a quantum model for the transport of energy in terms of quasiparticles, here referred to as
`solitons' \cite{Davydov1973,Davydov1976,Davydov1979,Davydov1982,Davydov1986,Davydov1987,Davydov1988},
where the full atomic complexity of the protein $\alpha$-helix
was reduced to the nonlinear interaction between amide~I vibrations
(excitons), and deformations (phonons) of the lattice of hydrogen bonds
that stabilize the helical structure \cite{Brizhik1983,Brizhik1988,Brizhik1993,Brizhik1995,Brizhik2004,Brizhik2006,Brizhik2010,Cruzeiro1988,Cruzeiro1994,Cruzeiro1997,Cruzeiro2009,Kerr1987,Kerr1990,MacNeil1984,Scott1984,Scott1985,Scott1992,Luo2011,Luo2017,Sun2010}.

In this work, we model only a single hydrogen-bonded spine in the protein $\alpha$-helix, rather than the 3-spine structure of the entire helix. In realistic protein $\alpha$-helices, the quantum dynamics of amide I energy in the 3-spine structure will also depend on inter-spine interactions and may give rise to complicated multihump solitons \cite{Hennig2002,Brizhik2019}, which the current study does not consider.
For a single $\alpha$-helix spine of hydrogen-bonded peptide groups,
the generalized Davydov Hamiltonian is a sum of three parts
\begin{equation}
\hat{H}=\hat{H}_{\textrm{ex}}+\hat{H}_{\textrm{ph}}+\hat{H}_{\textrm{int}}\label{eq:Hamiltonian}
\end{equation}
respectively for amide~I excitons $\hat{H}_{\textrm{ex}}$, hydrogen-bonded
lattice phonons $\hat{H}_{\textrm{ph}}$, and exciton--phonon interaction
$\hat{H}_{\textrm{int}}$. The three parts of the Hamiltonian include
only nearest neighbor interactions, and they are formally similar
to those featuring in Holstein polaron theory \cite{Holstein1959a,Holstein1959b,Cruzeiro1994}
\begin{alignat}{1}
\hat{H}_{\textrm{ex}} & =\sum_{n}\left[E_{0}\hat{a}_{n}^{\dagger}\hat{a}_{n}-J_{n+1}\hat{a}_{n}^{\dagger}\hat{a}_{n+1}-J_{n}\hat{a}_{n}^{\dagger}\hat{a}_{n-1}\right]\\
\hat{H}_{\textrm{ph}} & =\frac{1}{2}\sum_{n}\left[\frac{\hat{p}_{n}^{2}}{M_{n}}+w\left(\hat{u}_{n+1}-\hat{u}_{n}\right)^{2}\right]\\
\hat{H}_{\textrm{int}} & =\chi_{r}\sum_{n}\left(\hat{u}_{n+1}+\left(\xi-1\right)\hat{u}_{n}-\xi\hat{u}_{n-1}\right)\hat{a}_{n}^{\dagger}\hat{a}_{n}
\end{alignat}
where the index $n$ counts the peptide groups along the $\alpha$-helix
spine, $\hat{a}_{n}^{\dagger}$ and $\hat{a}_{n}$ are the boson creation
and annihilation operators for the amide~I excitons, $E_{0}$~is
the amide~I exciton energy, $J_{n}$~is the dipole--dipole coupling
energy between the $n$th and $(n-1)$th amide~I oscillator along
the spine, $\hat{p}_{n}$~is the momentum operator, $\hat{u}_{n}$
is the displacement operator from the equilibrium position of the
peptide group~$n$, $M_{n}$ is the mass of the peptide group $n$, 
$w$ is the spring constant of the hydrogen bonds in the lattice \cite{Davydov1976,Davydov1979,Scott1992},
$\chi_{r}$ and $\chi_{l}$ are anharmonic parameters arising
from the coupling between the amide~I exciton and the phonon lattice
displacements, respectively to the right or to the left, $\bar{\chi}=\frac{\chi_{r}+\chi_{l}}{2}$,
and $\xi=\frac{\chi_{l}}{\chi_{r}}$ is the anisotropy parameter of
the exciton--phonon interaction (by construction $\chi_{r}\neq0$,
and $0\leq\chi_{l}\leq\chi_{r}$ so that $\xi$ varies in the interval
$[0,1]$) \cite{Luo2017,GeorgievGlazebrook2019,GeorgievGlazebrook2019b}.

The quantum equations of motion for multi-quanta states of amide~I energy
can be derived from the Hamiltonian \eqref{eq:Hamiltonian} with the
use of the following generalized ansatz state vector \cite{Kerr1990}:
\begin{equation}
|\Psi(t)\rangle=|a(t)\rangle|b(t)\rangle=\frac{1}{\sqrt{Q!}}\left[\sum_{n}a_{n}(t)\hat{a}_{n}^{\dagger}\right]^{Q}|0_{\textrm{ex}}\rangle e^{-\frac{\imath}{\hbar}\sum_{j}\left(b_{j}(t)\hat{p}_{j}-c_{j}(t)\hat{u}_{j}\right)}|0_{\textrm{ph}}\rangle\label{eq:ansatz}
\end{equation}
where
\begin{align}
|a(t)\rangle & =\frac{1}{\sqrt{Q!}}\left[\sum_{n}a_{n}(t)\hat{a}_{n}^{\dagger}\right]^{Q}|0_{\textrm{ex}}\rangle\\
|b(t)\rangle & =e^{-\frac{\imath}{\hbar}\sum_{j}\left(b_{j}(t)\hat{p}_{j}-c_{j}(t)\hat{u}_{j}\right)}|0_{\textrm{ph}}\rangle
\end{align}
For ease of notation, the time dependence of $a_{n}(t)$,
$b_{n}(t)$, $c_{n}(t)$, $|\Psi(t)\rangle$, $|a(t)\rangle$ and
$|b(t)\rangle$, will henceforth be implicitly understood. Taking the inner product of
this generalized ansatz with itself gives
\begin{equation}
\langle\Psi|\Psi\rangle=\left[\sum_{n}|a_{n}|^{2}\right]^{Q}\label{eq:normalized}
\end{equation}
which implies
\begin{equation}
\sum_{n}|a_{n}|^{2}=1\label{eq:normalized-1}
\end{equation}
Verification of \eqref{eq:normalized} requires the multinomial theorem
\cite{Tauber1963}
\begin{equation}
\left[\sum_{n}a_{n}\right]^{Q}=\sum_{k_{1}+k_{2}+\ldots+k_{n}=Q}\left(\begin{array}{c}
Q\\
k_{1},k_{2},\ldots,k_{n}
\end{array}\right)a_{1}^{k_{1}}a_{2}^{k_{2}}\ldots a_{n}^{k_{n}}\label{eq:multinomial}
\end{equation}
where the multinomial coefficient is
\begin{equation}
\left(\begin{array}{c}
Q\\
k_{1},k_{2},\ldots,k_{n}
\end{array}\right)=\frac{Q!}{k_{1}!k_{2}!\ldots k_{n}!}
\end{equation}
with $k_{1}+k_{2}+\ldots+k_{n}=Q$. Then the inner product of the
ansatz is
\begin{equation}
\langle\Psi|\Psi\rangle=\frac{1}{Q!}\langle0_{\textrm{ex}}|\left[\sum_{n'}a_{n'}^{*}(t)\hat{a}_{n'}\right]^{Q}\left[\sum_{n}a_{n}(t)\hat{a}_{n}^{\dagger}\right]^{Q}|0_{\textrm{ex}}\rangle\langle b|b\rangle
\end{equation}
Taking into account that a non-zero result is only possible for creation
and annihilation operators with same powers at a given site, together
with $\langle b|b\rangle=1$ and
\begin{equation}
\langle0_{\textrm{ex}}|\left(a_{n}^{*}\hat{a}_{n}\right)^{k_{n}}\left(a_{n}\hat{a}_{n}^{\dagger}\right)^{k_{n}}|0_{\textrm{ex}}\rangle=\left(|a_{n}|^{2}\right)^{k_{n}}k_{n}!
\end{equation}
we obtain
\begin{align*}
\langle\Psi|\Psi\rangle & =\frac{1}{Q!}\sum_{k_{1}+k_{2}+\ldots+k_{n}=Q}\left(\frac{Q!}{k_{1}!k_{2}!\ldots k_{n}!}\right)^{2}\left(|a_{1}|^{2}\right)^{k_{1}}\left(|a_{2}|^{2}\right)^{k_{2}}\ldots\left(|a_{n}|^{2}\right)^{k_{n}}\\
 & \qquad\qquad\qquad\times\langle0_{\textrm{ex}}|\hat{a}_{1}^{k_{1}}\hat{a}_{2}^{k_{2}}\ldots\hat{a}_{n}^{k_{n}}\left(\hat{a}_{1}^{\dagger}\right)^{k_{1}}\left(\hat{a}_{2}^{\dagger}\right)^{k_{2}}\ldots\left(\hat{a}_{n}^{\dagger}\right)^{k_{n}}|0_{\textrm{ex}}\rangle\\
 & =\sum_{k_{1}+k_{2}+\ldots+k_{n}=Q}\frac{Q!}{k_{1}!k_{2}!\ldots k_{n}!}\left(|a_{1}|^{2}\right)^{k_{1}}\left(|a_{2}|^{2}\right)^{k_{2}}\ldots\left(|a_{n}|^{2}\right)^{k_{n}}=\left[\sum_{n}|a_{n}|^{2}\right]^{Q}
\end{align*}
For $Q=1$, the generalized ansatz $|\Psi(t)\rangle$ reduces to Davydov's
original $|D_{2}(t)\rangle$ ansatz \cite{Sun2010,Davydov1982}, which
was studied extensively in several previous works \cite{Kerr1987,GeorgievGlazebrook2019,GeorgievGlazebrook2019b}.

Given that the ansatz \eqref{eq:ansatz} is a suitable approximation to the exact
solution of the Schrödinger equation \cite{Dirac1967}, we obtain
\begin{equation}
\imath\hbar\frac{d}{dt}|\Psi\rangle=\hat{H}|\Psi\rangle\label{eq:schrodinger}
\end{equation}
For a given peptide group $n$, the expectation values for the exciton
number operator \mbox{$\hat{N}_{n}=\hat{a}_{n}^{\dagger}\hat{a}_{n}$}, phonon
displacement operator $\hat{u}_{n}$ and phonon momentum operator
$\hat{p}_{n}$ are
\begin{eqnarray}
\langle\hat{N}_{n}\rangle & = & \langle\Psi|\hat{N}_{n}|\Psi\rangle=Q\,|a_{n}|^{2}\label{eq:ave-a}\\
\langle\hat{u}_{n}\rangle & = & \langle\Psi|\hat{u}_{n}|\Psi\rangle=b_{n}\label{eq:ave-b}\\
\langle\hat{p}_{n}\rangle & = & \langle\Psi|\hat{p}_{n}|\Psi\rangle=c_{n}\label{eq:ave-c}
\end{eqnarray}
The verification of \eqref{eq:ave-a} requires the use of
\begin{equation}
\langle0_{\textrm{ex}}|\left(a_{n}^{*}\hat{a}_{n}\right)^{k_{n}}\hat{a}_{n}^{\dagger}\hat{a}_{n}\left(a_{n}\hat{a}_{n}^{\dagger}\right)^{k_{n}}|0_{\textrm{ex}}\rangle=\left(|a_{n}|^{2}\right)^{k_{n}}k_{n}k_{n}!
\end{equation}
Substitution of the ansatz \eqref{eq:ansatz} in $\langle\Psi|\hat{N}_{n}|\Psi\rangle$,
and using the multinomial theorem \eqref{eq:multinomial}, gives
\begin{align*}
\langle\Psi|\hat{N}_{n}|\Psi\rangle & =\frac{1}{Q!}\sum_{k_{1}+k_{2}+\ldots+k_{n}=Q}\left(\frac{Q!}{k_{1}!k_{2}!\ldots k_{n}!}\right)^{2}\left(|a_{1}|^{2}\right)^{k_{1}}\left(|a_{2}|^{2}\right)^{k_{2}}\ldots\left(|a_{n}|^{2}\right)^{k_{n}}k_{n}k_{1}!k_{2}!\ldots k_{n}!\\
 & =\sum_{k_{1}+k_{2}+\ldots+(k_{n}-1)=Q-1}\frac{Q(Q-1)!}{k_{1}!k_{2}!\ldots(k_{n}-1)!}\left(|a_{1}|^{2}\right)^{k_{1}}\left(|a_{2}|^{2}\right)^{k_{2}}\ldots\left(|a_{n}|^{2}\right)^{(k_{n}-1+1)}\\
 & =Q|a_{n}|^{2}\left[\sum_{n'}|a_{n'}|^{2}\right]^{Q-1}=Q|a_{n}|^{2}
\end{align*}
Verification of \eqref{eq:ave-b} requires the result that the expectation
values of the position and momentum operators of peptide groups with
the vacuum are zero, $\langle0_{\textrm{ph}}|\hat{u}_{n}|0_{\textrm{ph}}\rangle=0$
and $\langle0_{\textrm{ph}}|\hat{p}_{n}|0_{\textrm{ph}}\rangle=0$,
together with the Hadamard lemma:
\begin{eqnarray}
e^{\hat{A}}\hat{B}e^{-\hat{A}} & = & \exp\left(\textrm{ad}_{\hat{A}}\right)\left(\hat{B}\right)=\sum_{k=0}^{\infty}\frac{1}{k!}\left(\textrm{ad}_{\hat{A}}\right)^{k}\left(\hat{B}\right)\nonumber \\
 & = & \hat{B}+\left[\hat{A},\hat{B}\right]+\frac{1}{2!}\left[\hat{A},\left[\hat{A},\hat{B}\right]\right]+\frac{1}{3!}\left[\hat{A},\left[\hat{A},\left[\hat{A},\hat{B}\right]\right]\right]+\ldots\label{eq:hadamard}
\end{eqnarray}
where $\textrm{ad}_{\hat{A}}(\hat{B})\equiv\left[\hat{A},\hat{B}\right]$
is the adjoint operator. Substitution of the ansatz \eqref{eq:ansatz}
in $\langle\Psi|\hat{u}_{n}|\Psi\rangle$, and application of the Hadamard
lemma \eqref{eq:hadamard} with the standard quantum commutation relations
$[\hat{u}_{n},\hat{p}_{n}]=\imath\hbar$ and $[\hat{p}_{n},\hat{u}_{n}]=-\imath\hbar$,
gives
\begin{align*}
\langle\Psi|\hat{u}_{n}|\Psi\rangle & =\langle a|a\rangle\langle0_{\textrm{ph}}|e^{\frac{\imath}{\hbar}\sum_{j}\left(b_{j}\hat{p}_{j}-c_{j}\hat{u}_{j}\right)}\hat{u}_{n}e^{-\frac{\imath}{\hbar}\sum_{j}\left(b_{j}\hat{p}_{j}-c_{j}\hat{u}_{j}\right)}|0_{\textrm{ph}}\rangle\\
 & =\langle0_{\textrm{ph}}|e^{\frac{\imath}{\hbar}\left(b_{n}\hat{p}_{n}-c_{n}\hat{u}_{n}\right)}e^{\frac{\imath}{\hbar}\sum_{j\neq n}\left(b_{j}\hat{p}_{j}-c_{j}\hat{u}_{j}\right)}\hat{u}_{n}e^{-\frac{\imath}{\hbar}\sum_{j\neq n}\left(b_{j}\hat{p}_{j}-c_{j}\hat{u}_{j}\right)}e^{-\frac{\imath}{\hbar}\left(b_{n}\hat{p}_{n}-c_{n}\hat{u}_{n}\right)}|0_{\textrm{ph}}\rangle\\
 & =\langle0_{\textrm{ph}}|e^{\frac{\imath}{\hbar}\left(b_{n}\hat{p}_{n}-c_{n}\hat{u}_{n}\right)}e^{\frac{\imath}{\hbar}\sum_{j\neq n}\left(b_{j}\hat{p}_{j}-c_{j}\hat{u}_{j}\right)}e^{-\frac{\imath}{\hbar}\sum_{j\neq n}\left(b_{j}\hat{p}_{j}-c_{j}\hat{u}_{j}\right)}\hat{u}_{n}e^{-\frac{\imath}{\hbar}\left(b_{n}\hat{p}_{n}-c_{n}\hat{u}_{n}\right)}|0_{\textrm{ph}}\rangle\\
 & =\langle0_{\textrm{ph}}|e^{\frac{\imath}{\hbar}\left(b_{n}\hat{p}_{n}-c_{n}\hat{u}_{n}\right)}\hat{u}_{n}e^{-\frac{\imath}{\hbar}\left(b_{n}\hat{p}_{n}-c_{n}\hat{u}_{n}\right)}|0_{\textrm{ph}}\rangle\\
 & =\langle0_{\textrm{ph}}|\hat{u}_{n}+[\frac{\imath}{\hbar}\left(b_{n}\hat{p}_{n}-c_{n}\hat{u}_{n}\right),\hat{u}_{n}]|0_{\textrm{ph}}\rangle\\
 & =\langle0_{\textrm{ph}}|\hat{u}_{n}|0_{\textrm{ph}}\rangle+\langle0_{\textrm{ph}}|b_{n}|0_{\textrm{ph}}\rangle=b_{n}
\end{align*}
where we used $\langle a|a\rangle=1$ as implied from \eqref{eq:normalized}
and \eqref{eq:normalized-1}.
Verification of \eqref{eq:ave-c} is analogous \cite{GeorgievGlazebrook2019c} (only the last two steps are shown):
\begin{equation*}
\langle\Psi|\hat{p}_{n}|\Psi\rangle
  =\langle0_{\textrm{ph}}|\hat{p}_{n}+[\frac{\imath}{\hbar}\left(b_{n}\hat{p}_{n}-c_{n}\hat{u}_{n}\right),\hat{p}_{n}]|0_{\textrm{ph}}\rangle
  =\langle0_{\textrm{ph}}|\hat{p}_{n}|0_{\textrm{ph}}\rangle+\langle0_{\textrm{ph}}|c_{n}|0_{\textrm{ph}}\rangle=c_{n}
\end{equation*}

Identifying $b_{n}$ and $c_{n}$ as quantum expectation values
is important for a further application of the Schrödinger equation and
its complex conjugate in the form of the generalized Ehrenfest theorem
\cite{GeorgievGlazebrook2019}, which governs the quantum dynamics
of the expectation values
\begin{align}
\frac{d}{dt}b_{n} & =\frac{1}{\imath\hbar}\langle\Psi|\left[\hat{u}_{n},\hat{H}\right]|\Psi\rangle\label{eq:Ehrenfest-1}\\
\frac{d}{dt}c_{n} & =\frac{1}{\imath\hbar}\langle\Psi|\left[\hat{p}_{n},\hat{H}\right]|\Psi\rangle\label{eq:Ehrenfest-2}
\end{align}
From the Hamiltonian \eqref{eq:Hamiltonian}, together with $\left[\hat{u}_{n},\hat{p}_{n}^{2}\right]=2\imath\hbar\hat{p}_{n}$,
and $\left[\hat{p}_{n},\hat{u}_{n}^{2}\right]=-2\imath\hbar\hat{u}_{n}$,
we obtain the two commutators to be
\begin{align}
\left[\hat{u}_{n},\hat{H}\right] & =\imath\hbar\frac{\hat{p}_{n}}{M_{n}}\\
\left[\hat{p}_{n},\hat{H}\right] & =\imath\hbar w\Big(\hat{u}_{n-1}-2\hat{u}_{n}+\hat{u}_{n+1})-\imath\hbar\chi_{r}\Big(\hat{a}_{n-1}^{\dagger}\hat{a}_{n-1}+(\xi-1)\hat{a}_{n}^{\dagger}\hat{a}_{n}-\xi\hat{a}_{n+1}^{\dagger}\hat{a}_{n+1}\Big)
\end{align}
Substitution in \eqref{eq:Ehrenfest-1} and \eqref{eq:Ehrenfest-2},
followed by application of \eqref{eq:ave-a}, \eqref{eq:ave-b} and
\eqref{eq:ave-c}, yields one of Davydov's equations
\begin{equation}
M_{n}\frac{d^{2}}{dt^{2}}b_{n}=w\Big(b_{n-1}-2b_{n}+b_{n+1})-Q\chi_{r}\Big(\left|a_{n-1}\right|^{2}+(\xi-1)\left|a_{n}\right|^{2}-\xi\left|a_{n+1}\right|^{2}\Big)\label{eq:davydov-1}
\end{equation}

The equation for the amide~I probability amplitudes $a_{n}$ can be
derived by differentiating the $|\Psi\rangle$ ansatz using the product
rule
\begin{equation}
\imath\hbar\frac{d}{dt}|\Psi\rangle=\left(\imath\hbar\frac{d}{dt}|a\rangle\right)|b\rangle+|a\rangle\left(\imath\hbar\frac{d}{dt}|b\rangle\right)\label{eq:dPsi-dt}
\end{equation}
where (cf. \cite{Kerr1987,Kerr1990})
\begin{alignat}{1}
\imath\hbar\frac{d}{dt}|a\rangle & =\imath\hbar\sqrt{Q}\left(\sum_{n'}\frac{da_{n'}}{dt}\hat{a}_{n'}^{\dagger}\right)\frac{1}{\sqrt{(Q-1)!}}\left(\sum_{n}a_{n}\hat{a}_{n}^{\dagger}\right)^{Q-1}|0_{\textrm{ex}}\rangle\\
\imath\hbar\frac{d}{dt}|b\rangle & =\sum_{n}\left[\frac{db_{n}}{dt}\hat{p}_{n}-\frac{dc_{n}}{dt}\hat{u}_{n}+\frac{1}{2}\left(b_{n}\frac{dc_{n}}{dt}-\frac{db_{n}}{dt}c_{n}\right)\right]|b\rangle
\end{alignat}

Next, we use the Schrödinger equation \eqref{eq:schrodinger} and
take the inner product with the state $\frac{1}{\sqrt{Q!}}\langle b|\langle0_{\textrm{ex}}|\left(\hat{a}_{n}\right)^{Q}$
as follows:
\begin{equation}
\frac{1}{\sqrt{Q!}}\langle b|\langle0_{\textrm{ex}}|\left(\hat{a}_{n}\right)^{Q}\imath\hbar\frac{d}{dt}|\Psi\rangle=\frac{1}{\sqrt{Q!}}\langle b|\langle0_{\textrm{ex}}|\left(\hat{a}_{n}\right)^{Q}\hat{H}|\Psi\rangle\label{eq:inner-Q}
\end{equation}
Again, a non-zero result is only possible if the creation and annihilation
operators at a given site have the same power. Since the operators
act at most on two different sites, the use of the multinomial theorem
reduces to the binomial one (there are at most two sites with non-zero
powers). Straightforward application of binomial coefficients ${n \choose k}=\frac{n!}{k!(n-k)!}$
allows the calculation of individual inner products
\begin{align*}
\frac{1}{\sqrt{Q!}}\langle0_{\textrm{ex}}|\left(\hat{a}_{n}\right)^{Q}\frac{d}{dt}|a\rangle & =\frac{1}{\sqrt{Q!}}\frac{Q}{\sqrt{Q}}\frac{da_{n}}{dt}\frac{1}{\sqrt{(Q-1)!}}{Q-1 \choose Q-1}a_{n}^{Q-1}\langle0_{\textrm{ex}}|\left(\hat{a}_{n}\right)^{Q}\left(\hat{a}_{n}^{\dagger}\right)^{Q}|0_{\textrm{ex}}\rangle\\
 & =\frac{da_{n}}{dt}a_{n}^{Q-1}Q
\end{align*}
\begin{align*}
\frac{1}{\sqrt{Q!}}\langle0_{\textrm{ex}}|\left(\hat{a}_{n}\right)^{Q}|a\rangle & =\frac{1}{\sqrt{Q!}}\frac{1}{\sqrt{Q!}}{Q \choose Q}a_{n}^{Q}\langle0_{\textrm{ex}}|\left(\hat{a}_{n}\right)^{Q}\left(\hat{a}_{n}^{\dagger}\right)^{Q}|0_{\textrm{ex}}\rangle\\
 & =a_{n}^{Q}
\end{align*}
\begin{align*}
\frac{1}{\sqrt{Q!}}\langle0_{\textrm{ex}}|\left(\hat{a}_{n}\right)^{Q}\hat{a}_{n}^{\dagger}\hat{a}_{n}|a\rangle & =\frac{1}{\sqrt{Q!}}\frac{1}{\sqrt{Q!}}{Q \choose Q}a_{n}^{Q}\langle0_{\textrm{ex}}|\left(\hat{a}_{n}\right)^{Q}\hat{a}_{n}^{\dagger}\hat{a}_{n}\left(\hat{a}_{n}^{\dagger}\right)^{Q}|0_{\textrm{ex}}\rangle\\
 & =a_{n}^{Q}Q
\end{align*}
\begin{align*}
\frac{1}{\sqrt{Q!}}\langle0_{\textrm{ex}}|\left(\hat{a}_{n}\right)^{Q}\hat{a}_{n}^{\dagger}\hat{a}_{n\pm1}|a\rangle & =\frac{1}{\sqrt{Q!}}\frac{1}{\sqrt{Q!}}{Q \choose Q-1}a_{n\pm1}a_{n}^{Q-1}\langle0_{\textrm{ex}}|\left(\hat{a}_{n}\right)^{Q}\left(\hat{a}_{n}^{\dagger}\right)^{Q}|0_{\textrm{ex}}\rangle\\
 & =a_{n\pm1}a_{n}^{Q-1}Q
\end{align*}
Substitution of \eqref{eq:dPsi-dt} together with the Hamiltonian
\eqref{eq:Hamiltonian} into \eqref{eq:inner-Q}, together with the
expectation values \eqref{eq:ave-b} and \eqref{eq:ave-c}, after
cancellation of common factor $a_{n}^{Q-1}Q$ yields Davydov's second
equation
\begin{equation}
\imath\hbar\frac{d}{dt}a_{n}=\gamma(t)a_{n}-J_{n+1}a_{n+1}-J_{n}a_{n-1}+\chi_{r}\left(b_{n+1}+\left(\xi-1\right)b_{n}-\xi b_{n-1}\right)a_{n}\label{eq:second-raw}
\end{equation}
where $\gamma(t)$ is a real-valued global term for all sites $n$
given by
\begin{equation}
\gamma(t)=E_{0}+\frac{1}{Q}\left[W(t)+\frac{1}{2}\sum_{j}\left(b_{j}\frac{dc_{j}}{dt}-\frac{db_{j}}{dt}c_{j}\right)\right]
\end{equation}
The phonon energy $W(t)$ can be computed with the use of the Hadamard
lemma \eqref{eq:hadamard} again utilizing the zero expectation values
of the position and momentum operators of peptide groups with the
vacuum, together with $c_{j}=M_{j}\frac{db_{j}}{dt}$, and the following
commutators
\begin{gather*}
[-\frac{\imath}{\hbar}c_{j}\hat{u}_{j},\frac{1}{2}\frac{\hat{p}_{j}^{2}}{M_{j}}]=\frac{c_{j}}{M_{j}}\hat{p}_{j}\\{}
[-\frac{\imath}{\hbar}c_{j}\hat{u}_{j},\frac{c_{j}}{M_{j}}\hat{p}_{j}]=\frac{c_{j}^{2}}{M_{j}}\\{}
[\frac{\imath}{\hbar}\left(b_{j}\hat{p}_{j}+b_{j+1}\hat{p}_{j+1}\right),\frac{1}{2}w\left(\hat{u}_{j+1}^{2}-2\hat{u}_{j}\hat{u}_{j+1}+\hat{u}_{j}^{2}\right)]=w\left(b_{j+1}\hat{u}_{j+1}-b_{j}\hat{u}_{j+1}-b_{j+1}\hat{u}_{j}+b_{j}\hat{u}_{j}\right)\\{}
[\frac{\imath}{\hbar}\left(b_{j}\hat{p}_{j}+b_{j+1}\hat{p}_{j+1}\right),w\left(b_{j+1}\hat{u}_{j+1}-b_{j}\hat{u}_{j+1}-b_{j+1}\hat{u}_{j}+b_{j}\hat{u}_{j}\right)]=w\left(b_{j+1}^{2}-2b_{j}b_{j+1}+b_{j}^{2}\right)
\end{gather*}
\begin{equation}
W(t)=\langle b|H_{\textrm{ph}}|b\rangle=W_{0}+\frac{1}{2}\sum_{j}\left[M_{j}\left(\frac{db_{j}}{dt}\right)^{2}+w\left(b_{j+1}-b_{j}\right)^{2}\right]
\end{equation}
with $W_{0}=\langle0_{\textrm{ph}}|\hat{H}_{\textrm{ph}}|0_{\textrm{ph}}\rangle$
denoting the zero-point phonon energy.

In order to make further headway, we note that introduction of a global phase change
on the exciton quantum probability amplitudes, namely $a_{n}\to\bar{a}_{n}e^{-\frac{\imath}{\hbar}\int\gamma(t)dt}$,
will not change the quantum probabilities for finding the exciton
at a given site
\begin{equation}
|a_{n}|^{2}=e^{+\frac{\imath}{\hbar}\int\gamma(t)dt}\bar{a}_{n}^{*}\bar{a}_{n}e^{-\frac{\imath}{\hbar}\int\gamma(t)dt}
\end{equation}
This is important because we will not need to transform the first
Davydov equation \eqref{eq:davydov-1}. After differentiation of the
left-hand side of \eqref{eq:second-raw}
\[
\imath\hbar\frac{d}{dt}\left(\bar{a}_{n}e^{-\frac{\imath}{\hbar}\int\gamma(t)dt}\right)=\imath\hbar e^{-\frac{\imath}{\hbar}\int\gamma(t)dt}\frac{d\bar{a}_{n}}{dt}+\bar{a}_{n}\gamma(t)e^{-\frac{\imath}{\hbar}\int\gamma(t)dt}
\]
followed by cancellation of common terms and relabeling of $\bar{a}_{n}$
back to $a_{n}$, we obtain the following system of gauge transformed
quantum equations of motion
\begin{eqnarray}
\imath\hbar\frac{d}{dt}a_{n} & = & -J_{n+1}a_{n+1}-J_{n}a_{n-1}+\chi_{r}\left[b_{n+1}+(\xi-1)b_{n}-\xi b_{n-1}\right]a_{n}\label{eq:gauge-1}\\
M_{n}\frac{d^{2}}{dt^{2}}b_{n} & = & w\Big(b_{n-1}-2b_{n}+b_{n+1})-Q\chi_{r}\Big(\left|a_{n-1}\right|^{2}+(\xi-1)\left|a_{n}\right|^{2}-\xi\left|a_{n+1}\right|^{2}\Big)\label{eq:gauge-2}
\end{eqnarray}

\section{A computational study}

\subsection{Model parameters}

The system of generalized Davydov equations \eqref{eq:gauge-1} and
\eqref{eq:gauge-2} for multiple amide~I quanta expands the scope
of the original Davydov model (cf. \cite{Brizhik1983,Brizhik1988,Brizhik1993,Brizhik2004,Brizhik2006,Brizhik2010,Davydov1976,Davydov1979,Davydov1982,Davydov1986,Davydov1987,Davydov1988,Kerr1987,Kerr1990,Kivshar1989,GeorgievGlazebrook2019,GeorgievGlazebrook2019b,Forner1990,Forner1991c})
to the extent that it is now capable of answering a number of questions
in regard to the quantum boundaries of life. In particular, anisotropy
of the exciton--phonon coupling for different values of the parameter
$\xi$, nonuniformity of amino acid masses $M_{n}$, and non--uniformity
of amide~I dipole--dipole coupling energies $J_{n}$ could be directly
incorporated in computer simulations and the resulting effects on
quantum dynamics could be theoretically characterized. In order to improve on the comparison
with a number of previous studies \cite{Cruzeiro1988,MacNeil1984,Scott1984,Scott1985,GeorgievGlazebrook2019,GeorgievGlazebrook2019b},
we here apply the following basic model parameters: spring constant
of the hydrogen bonds in the lattice $w=13$~N/m \cite{Itoh1972},
anharmonic parameter for the exciton--phonon coupling $\bar{\chi}=35$~pN
\cite{GeorgievGlazebrook2019}, average mass of an amino acid inside
the protein $\alpha$-helix $M=1.9\times10^{-25}$~kg \cite{Cruzeiro1988},
average amide~I dipole--dipole coupling energy $J=0.155$ zJ \cite{Nevskaya1976},
initially unperturbed lattice of hydrogen bonds, and $Q$ non-overlapping
Gaussian pulses of amide~I energy each spread over 5 peptide groups
with quantum probability amplitudes $a_{n}$ given by $\sqrt{\frac{1}{Q}}\{\sqrt{0.099},\sqrt{0.24},\sqrt{0.322},\sqrt{0.24},\sqrt{0.099}\}$. For simulations with a double soliton, we used $Q-1$ non-overlapping
Gaussian pulses, of which the double soliton had a factor of $\sqrt{\frac{2}{Q}}$ whereas the single solitons had a factor of $\sqrt{\frac{1}{Q}}$.
For most of the simulations, we used an $\alpha$-helix spine with length of $n_{\max}=40$ peptide groups, which covers a distance of 18~nm.

\subsection{\label{sub:4-2}Solitons in short protein $\alpha$-helices}

Davydov's original analysis aimed to establish the existence of
soliton (quasiparticle) solutions via derivation of nonlinear Schrödinger equation
(NLSE). To achieve that goal, the exciton--phonon interaction was
taken to be completely isotropic $\xi=1$, and all $J_{n}$ and $M_{n}$
were uniformly replaced with the corresponding average values $J$
and $M$. Davydov further introduced the dimensionless variable $x=\frac{\tilde{x}}{a}$,
where $\tilde{x}$ is the distance and $a$ is the spacing between
peptide groups \cite{Davydov1976}. He then approximated the system
of discrete functions with continuous ones using the transformation
$f(x\pm1,t)\approx(1\pm\frac{\partial}{\partial x}+\frac{1}{2}\frac{\partial^{2}}{\partial x^{2}})f(x,t)$
\cite{Davydov1979}. Lastly, the resulting system of partial differential
equations (PDEs) was manipulated into NLSE, which is known to have
analytic soliton solutions \cite{Kivshar1989}. However, because a
large number of mathematical assumptions enter at each step of Davydov's
derivation, it is not known whether the soliton will persist in short
protein $\alpha$-helices in the presence of anisotropy and variability
of various parameters.

\begin{figure}[t]
\begin{centering}
\includegraphics[width=129mm]{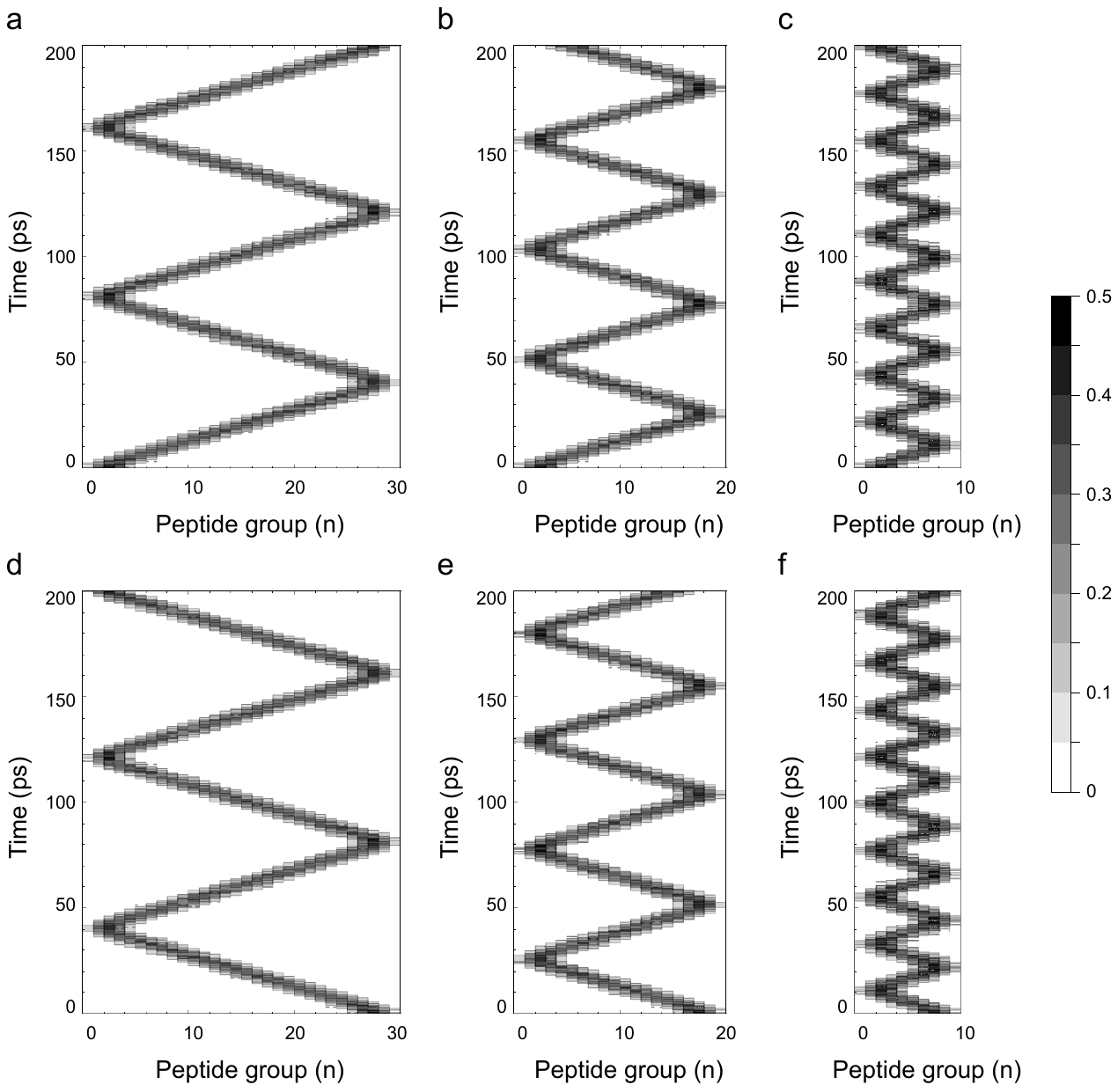}
\par\end{centering}

\caption{\label{fig:2}The quantum dynamics of a moving Davydov soliton for completely
isotropic exciton--phonon interaction $\xi=1$, with $Q=1$ quantum
of amide~I energy visualized through the expectation value of the
exciton number operator $Q|a_{n}|^{2}$ at each peptide group $n$.
(a--c) Gaussian pulse of amide~I energy is applied at the N-end of
a short protein $\alpha$-helix spine with varying length, $n_{\max}=30$
(a), $n_{\max}=20$ (b) or $n_{\max}=10$ (c). (d--f) Gaussian pulse
of amide~I energy is applied at the C-end of a short protein $\alpha$-helix
spine with varying length, $n_{\max}=30$ (d), $n_{\max}=20$ (e)
or $n_{\max}=10$ (f).}
\end{figure}

In order to systematize the effect of the protein $\alpha$-helix length
upon these solitons, we have integrated numerically the system of
Davydov equations with $Q=1$ for completely isotropic exciton--phonon
interaction $\xi=1$ and uniform values $J$ and $M$ for all peptide
groups $n$. The quantum dynamics of solitons launched from
either the N-end or the C-end of the $\alpha$-helix exhibited left--right
mirror symmetry (Fig.~\ref{fig:2}). The solitons moved with velocity
of 334 m/s, which is slower than 1 peptide group per picosecond (PG/ps), 1~PG/ps = 450~m/s.
The solitons were capable of multiple reflections from the $\alpha$-helix
ends without disintegration. In fact, discreteness effects did not
manifest even for ultrashort helices whose length was only $n_{\max}=10$
peptide groups (Figs.~\ref{fig:2}c,f), even though the initial soliton
width was already $5$ peptide groups, which is half the length of
the helix. This reveals that Davydov's intuition in regard to the approximations,
was essentially correct; the continuum approximation is valid for
$\xi=1$ regardless of the actual protein $\alpha$-helix length and
the soliton width.

\begin{figure}[t]
\begin{centering}
\includegraphics[width=129mm]{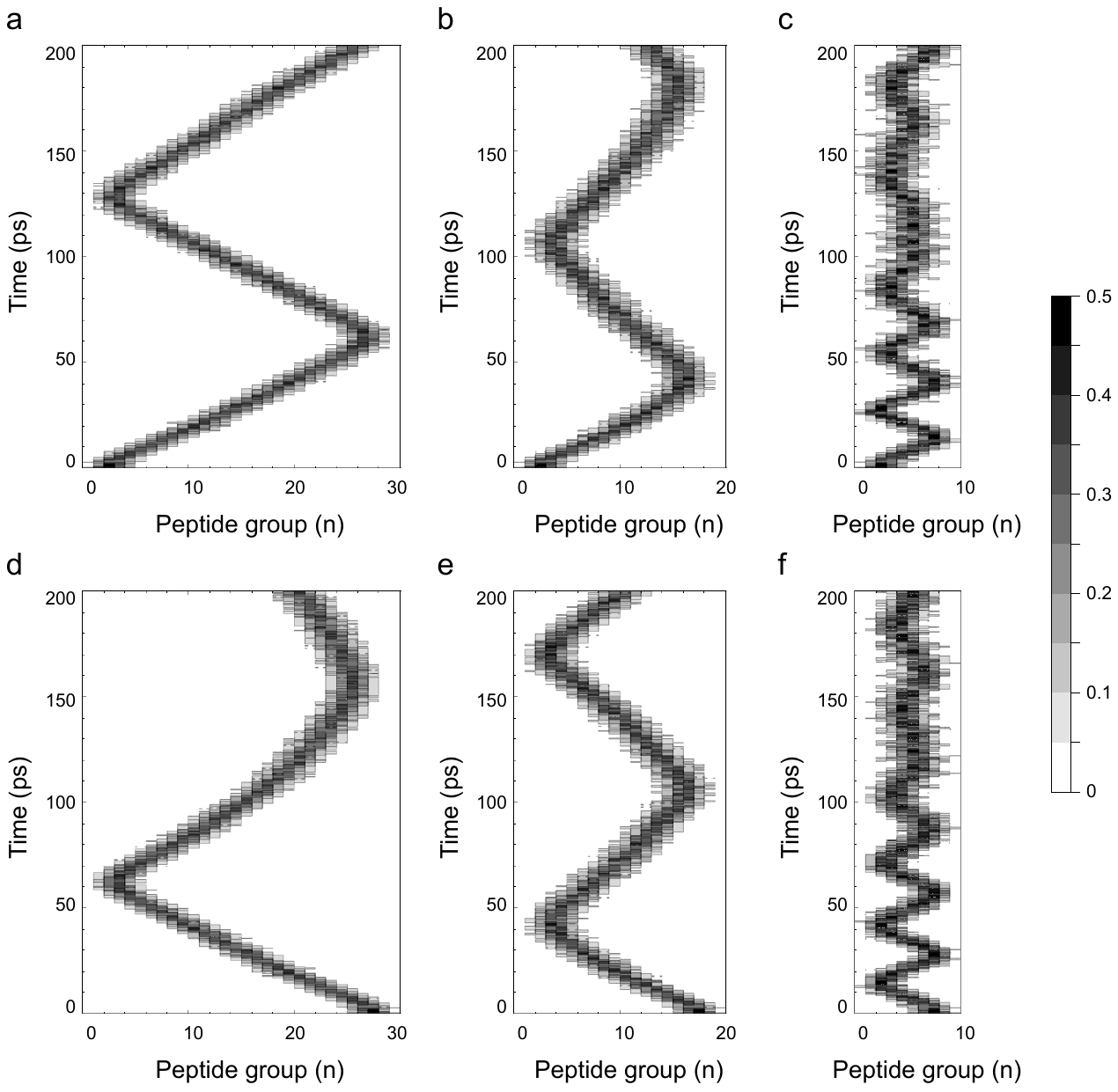}
\par\end{centering}

\caption{\label{fig:3} The quantum dynamics of a moving Davydov soliton for completely
anisotropic exciton--phonon interaction $\xi=0$, with $Q=1$ quantum
of amide~I energy visualized through the expectation value of the
exciton number operator $Q|a_{n}|^{2}$ at each peptide group $n$.
(a--c) The Gaussian pulse of amide~I energy is applied at the N-end of
a short protein $\alpha$-helix spine with varying length, $n_{\max}=30$
(a), $n_{\max}=20$ (b) or $n_{\max}=10$ (c). (d--f) The Gaussian pulse
of amide~I energy is applied at the C-end of a short protein $\alpha$-helix
spine with varying length, $n_{\max}=30$ (d), $n_{\max}=20$ (e)
or $n_{\max}=10$ (f).}
\end{figure}

Next, we have integrated numerically the system of Davydov equations
for completely anisotropic exciton--phonon interaction $\xi=0$, which
coincides with the model extensively studied by Scott and collaborators
\cite{Scott1984,Scott1985,Scott1992,Cruzeiro1988}. The speed of launched
solitons was lower 218 m/s and wobbled in time (Fig.~\ref{fig:3}).
In fact, for ultrashort helices whose length was only $n_{\max}=10$
peptide groups, the soliton reflected 6-7 times before settling in
the middle of the helix within a timescale of 100 ps (Figs.~\ref{fig:3}c,f).
Thus, the analytic soliton solution that moves at a constant speed
fails to capture correctly the quantum dynamics of the system for
$\xi=0$.

\begin{figure}[t]
\begin{centering}
\includegraphics[width=129mm]{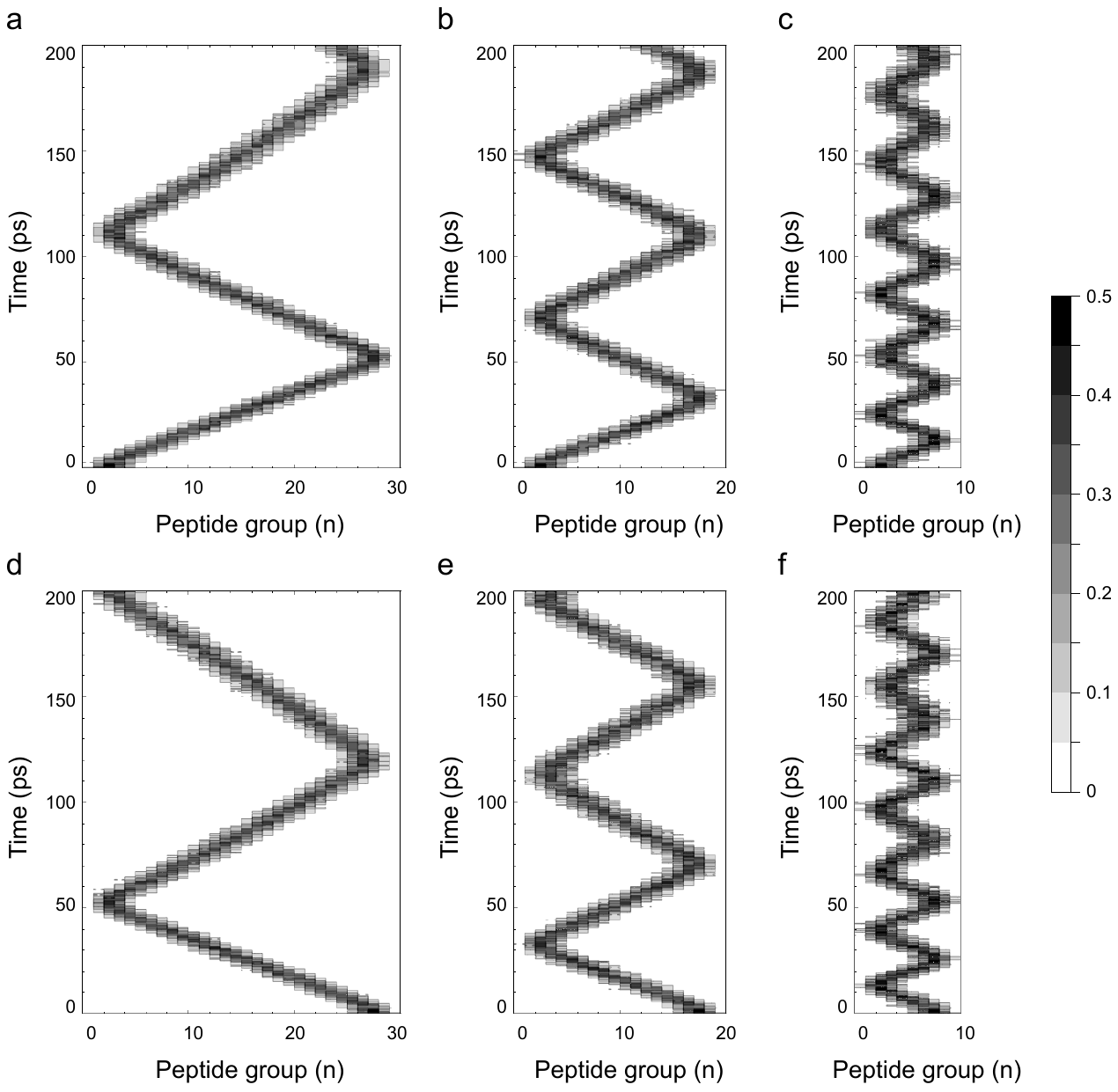}
\par\end{centering}

\caption{\label{fig:4}The quantum dynamics of a moving Davydov soliton for very
slightly isotropic exciton--phonon interaction $\xi=0.02$, with $Q=1$
quantum of amide~I energy visualized through the expectation value
of the exciton number operator $Q|a_{n}|^{2}$ at each peptide group~$n$.
(a--c) Gaussian pulse of amide~I energy is applied at the N-end
of a short protein $\alpha$-helix spine with varying length, $n_{\max}=30$
(a), $n_{\max}=20$ (b) or $n_{\max}=10$ (c). (d--f) Gaussian pulse
of amide~I energy is applied at the C-end of a short protein $\alpha$-helix
spine with varying length, $n_{\max}=30$ (d), $n_{\max}=20$ (e)
or $n_{\max}=10$ (f).}
\end{figure}

In actual protein $\alpha$-helices, the case of complete anisotropy of exciton--phonon
interaction, $\xi=\frac{\chi_{l}}{\chi_{r}}=0$, cannot be reached
as this will require absolutely no interaction of the amide~I oscillator
with the hydrogen bond to the left ($\chi_{l}=0$). From a theoretical
perspective, however, letting $\xi=0$ provides a limiting case that
may exhibit qualitatively different quantum behavior. Thus, varying
$\xi$ within the interval $[0,1]$ in systematic steps could provide
a threshold for qualitative change in quantum dynamics. Interestingly,
in the current context, such a threshold value exists and this could be
set to $\xi=0.02$. The presence of very slight isotropy of the exciton--phonon
interaction $\xi=0.02$ was able to restore the linear motion of the
soliton increasing its speed to 241 m/s (Fig.~\ref{fig:4}). The
soliton reflected repeatedly in ultrashort helices with $n_{\max}=10$
peptide groups for the whole time period of 200 ps of the simulation
(Figs.~\ref{fig:4}c,f), similarly to the completely isotropic case
with $\xi=1$. In other words, for realistic protein $\alpha$-helices
the main effect of the anisotropy of exciton--phonon interaction could
be to reduce the velocity of propagation.

Collectively, the above results indicate that the Davydov solitons are
robust quasiparticles with respect to lattice discreteness, and in turn, with respect
to the total length of the protein $\alpha$-helix. Thus, in the process
of evolutionary design, and optimization of protein functions in general, it is
physically plausible that the lengths of various protein $\alpha$-helices
in fact gradually evolved through steps of a single amino acid residue towards
achieving a certain optimal length for the delivery of free energy at
a desired active site. For the remainder of the simulations, in order to ensure a
sufficiently large arena with high spatial resolution of multi-soliton
dynamics with $Q>1$, we will consider a protein $\alpha$-helix spine
with a fixed length of $n_{\max}=40$ peptide groups. This length
has been also used in our previous works \cite{GeorgievGlazebrook2019,GeorgievGlazebrook2019b}
and allows direct comparison of the newly obtained results with the data that has already been 
published.

\subsection{\label{sub:4-3}Multi-quanta solitons}

\begin{figure}[t]
\begin{centering}
\includegraphics[width=130mm]{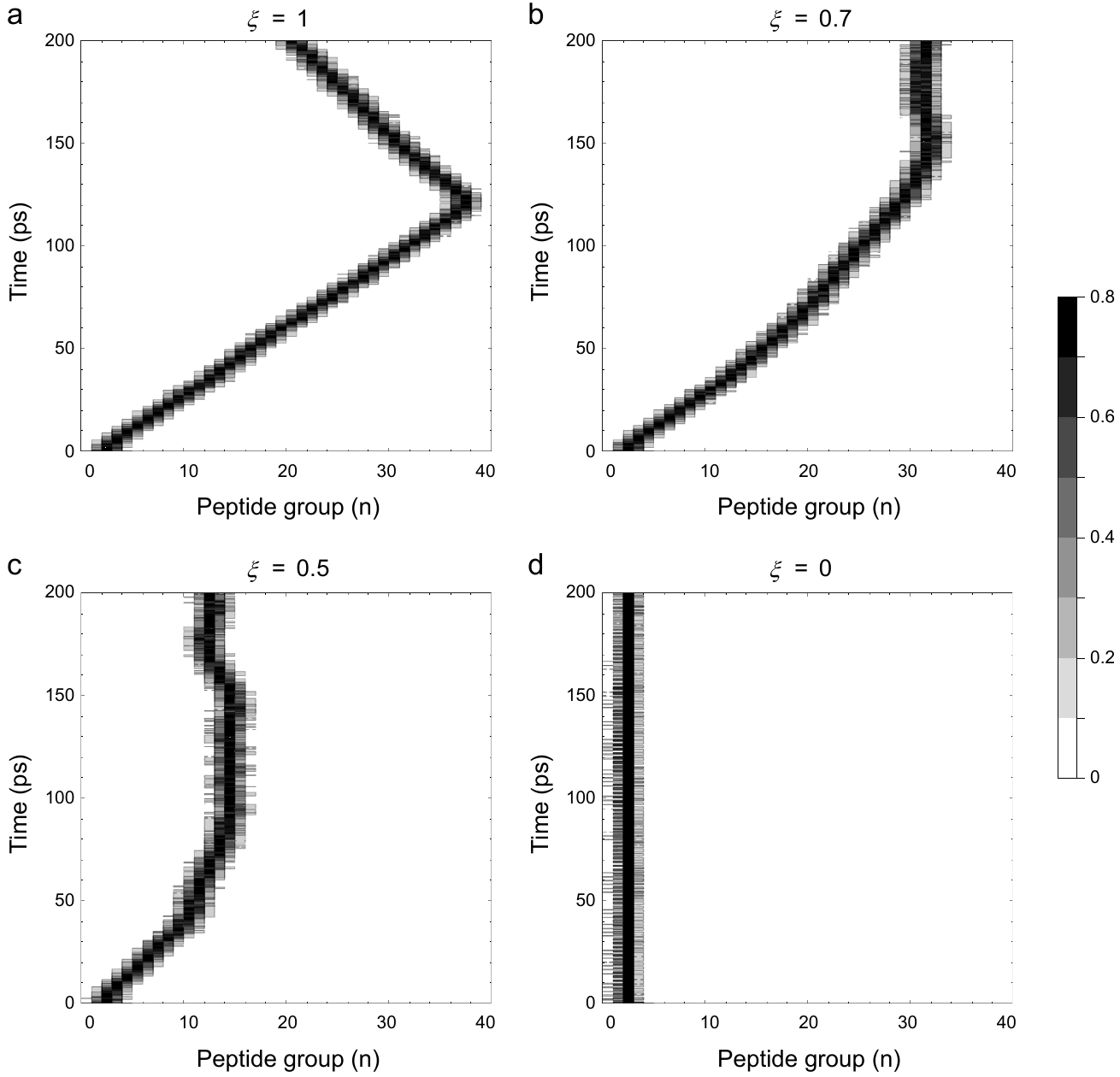}
\par\end{centering}

\caption{\label{fig:5}The quantum dynamics of a double Davydov soliton with $Q=2$
quanta of amide~I energy delivered as a single Gaussian pulse over
5 peptide groups at the N-end of a protein $\alpha$-helix visualized
through the expectation value of the exciton number operator $Q|a_{n}|^{2}$.
Decreasing the isotropy of exciton--phonon interaction leads to reduction
in soliton velocity and eventual soliton pinning: $\xi=1$ (a), $\xi=0.7$
(b), $\xi=0.5$ (c) and $\xi=0$ (d).}
\end{figure}

Previously we have shown that for $Q=1$, pulses of Gaussian amide
I energy are able to launch traveling solitons when applied to the protein $\alpha$-helix
ends, but generate pinned solitons if the amide~I energy is applied
in the interior of the protein $\alpha$-helix \cite{GeorgievGlazebrook2019}.
To study the collision of several Davydov solitons for $Q>1$, here
we have considered a number of different scenarios in which the anisotropy
of exciton--phonon interaction was systematically varied from $\xi=1$
to $\xi=0$ in steps of $0.1$. Because the quantum dynamics was affected 
nonlinearly with variation of $\xi$, we have presented
four panels per simulation in such a way that the exhibited changes
between any two consecutive panels are the most prominent.

\subsubsection{Double solitons}

\begin{figure}[t]
\begin{centering}
\includegraphics[width=130mm]{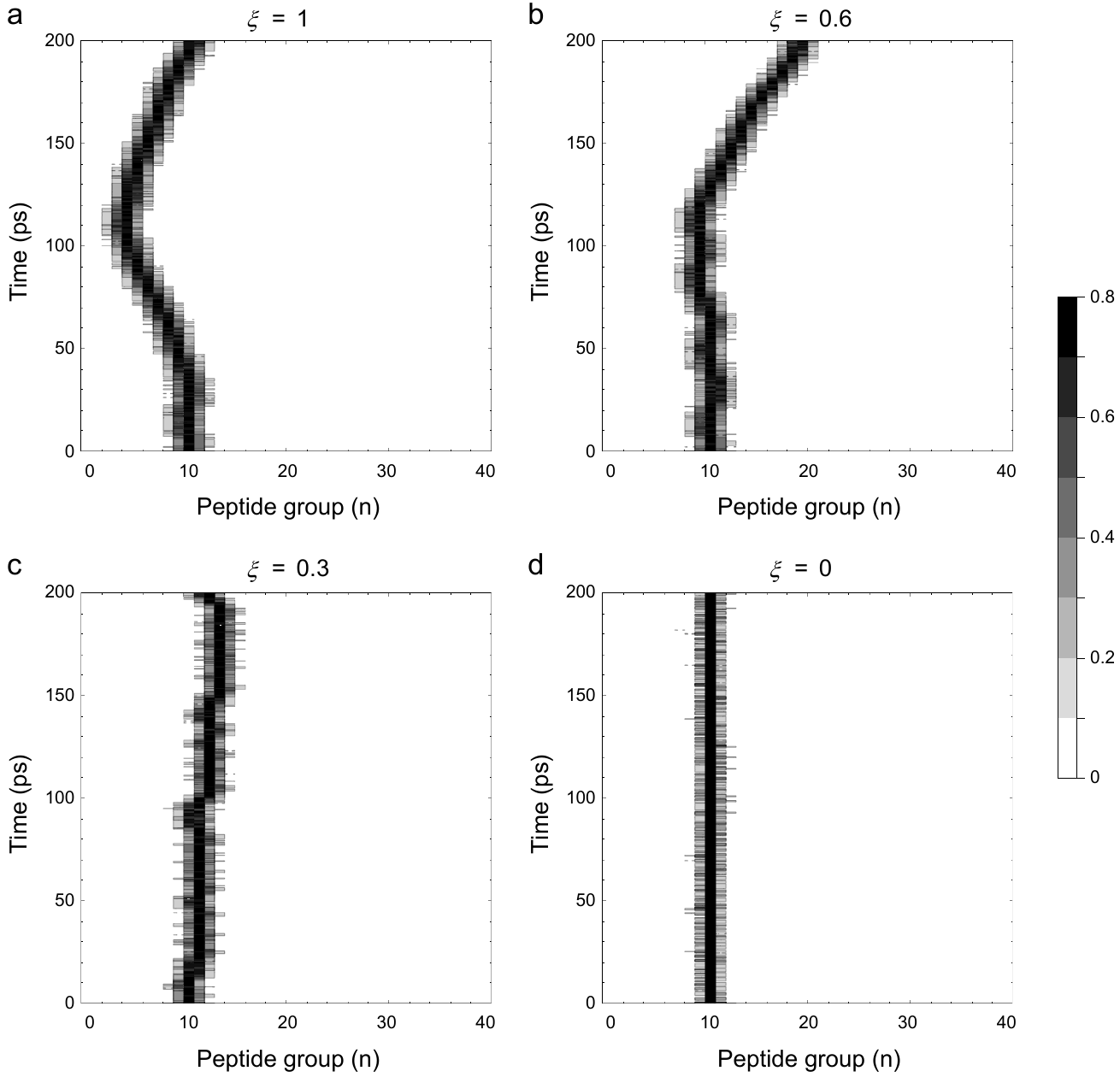}
\par\end{centering}

\caption{\label{fig:6}The quantum dynamics of a double Davydov soliton with $Q=2$
quanta of amide~I energy delivered as a single Gaussian pulse over
5 peptide groups $n=9-13$ in the interior of a protein $\alpha$-helix
visualized through the expectation value of the exciton number operator
$Q|a_{n}|^{2}$. Decreasing the isotropy of exciton--phonon interaction
leads to reduction in wobbling of the pinned soliton: $\xi=1$ (a),
$\xi=0.6$ (b), $\xi=0.3$ (c) and $\xi=0$ (d).}
\end{figure}

The simplest extension of the $Q=1$ case is to apply a single Gaussian
of amide~I energy over 5 peptide groups while exciting $Q=2$ amide
I quanta. As noted by Kerr and Lomdahl \cite{Kerr1990}, the multi-quantum
property of the ansatz state \eqref{eq:ansatz} results in a stronger
driving force on the phonon modes \eqref{eq:gauge-2}, but no modification
of the equation for the amide~I exciton probability amplitudes \eqref{eq:gauge-1}.
Thus, it might be expected that double solitons may exhibit features
similar to those resulting from an increased exciton--phonon coupling
$\bar{\chi}$; namely, greater soliton stability, lower soliton velocity,
and assisted soliton pinning \cite{GeorgievGlazebrook2019}. Effectively,
double Davydov solitons with $Q=2$ launched from the N-end of the
protein $\alpha$-helix for completely isotropic exciton--phonon interaction
$\xi=1$ moved at lower velocity of 147 m/s (Fig.~\ref{fig:5}a)
compared with velocity of 334 m/s for $Q=1$ (Fig.~\ref{fig:2}a).
Similarly, for completely anisotropic exciton--phonon interaction
$\xi=0$ the velocity of the soliton with $Q=2$ was also lower, in
fact zero due to soliton pinning (Fig.~\ref{fig:5}d), compared with
velocity of 218 m/s for $Q=1$ (Fig. \ref{fig:3}a). For intermediate
values of $\xi$, the soliton migrated towards the interior of the
protein $\alpha$-helix where it started wobbling around some fixed
interior point (Figs.~\ref{fig:5}b,c).

When the double soliton was generated in the interior of the protein
$\alpha$-helix, it was pinned and wobbled around its initial position
(Fig.~\ref{fig:6}). When the isotropy of exciton--phonon interaction
$\xi$ was decreased, this led to reduction in the wobbling of the
pinned soliton, which eventually came to a complete halt for $\xi=0$ (Fig.
\ref{fig:6}d). Thus, increasing the number of amide~I quanta increases
the energy of the soliton, enhances soliton stability and in turn lowers the
soliton propagation velocity.

\subsubsection{Two-soliton collisions}

\begin{figure}[t]
\begin{centering}
\includegraphics[width=130mm]{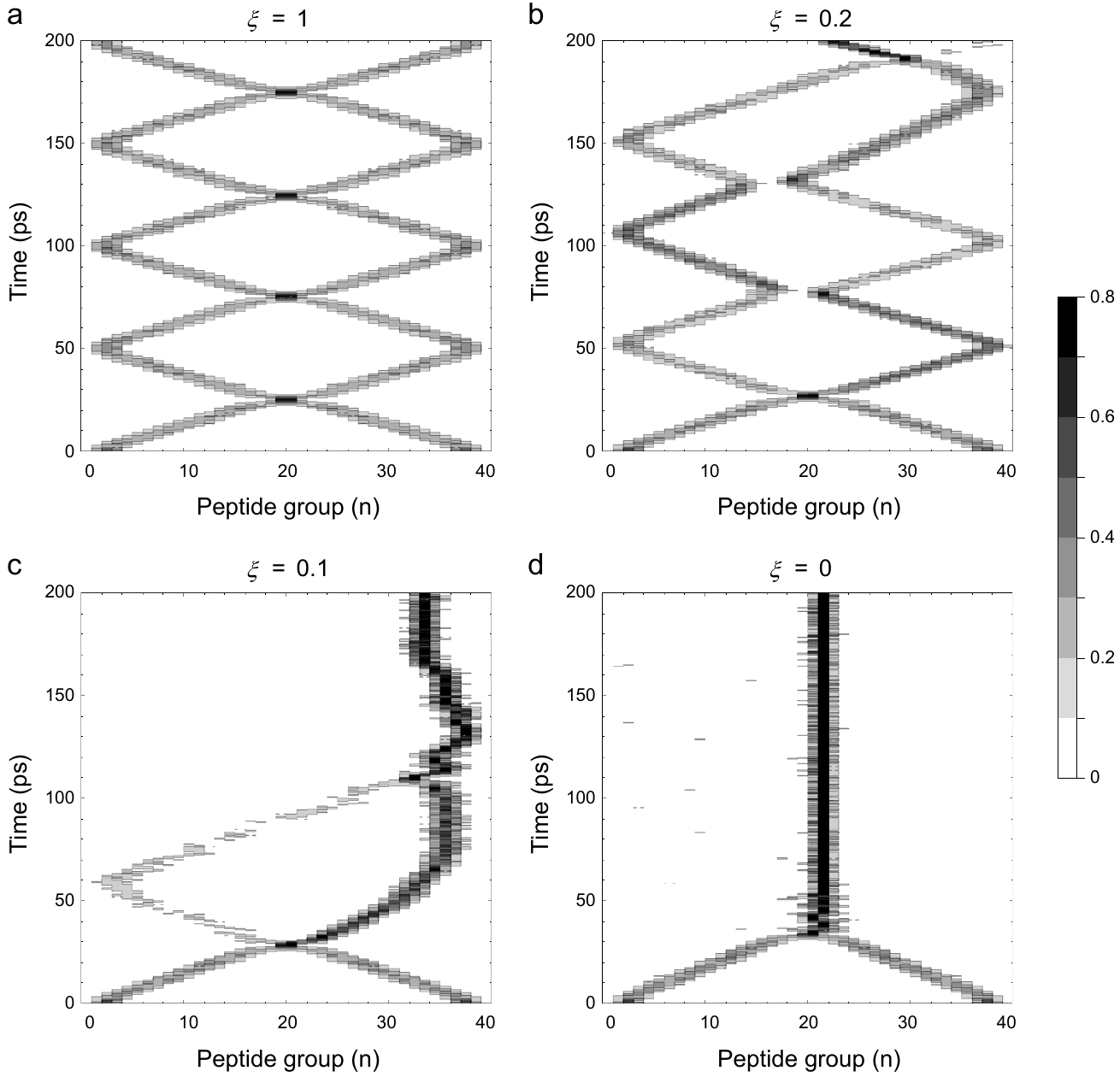}
\par\end{centering}

\caption{\label{fig:7} The quantum collision of two moving Davydov solitons with
$Q=2$ quanta of amide~I energy launched from the two ends of a protein
$\alpha$-helix visualized through the expectation value of the exciton
number operator $Q|a_{n}|^{2}$. Constructive quantum interference
may focus the amide~I energy at the collision site to a width much
narrower than each of the individual solitons. Decreasing the isotropy
of exciton--phonon interaction may result in soliton pinning after
the collision: $\xi=1$ (a), $\xi=0.2$ (b), $\xi=0.1$ (c) and $\xi=0$
(d).}
\end{figure}

The presence of multiple amide~I quanta in the protein $\alpha$-helix
allows for the application of an initial multi-Gaussian distribution,
which is a sum of several Gaussians, in order to generate several
solitons whose eventual collision may lead to detecting either constructive,
or destructive, quantum interference phenomena. This could be instrumental
in how proteins utilize free energy for driving life-supporting bio-processes.

The launching of two propagating solitons from the two ends of the protein
$\alpha$-helix, for $\xi=1$, leads to significant constructive quantum
interference of amide~I quantum probability amplitudes $a_{n}$, which
focuses the amide~I energy at the collision site to a width much narrower
than each of the individual solitons (Fig.~\ref{fig:7}a). Thus,
constructive quantum interference may provide a mechanism for the brief
focusing of energy at protein active centers for catalysis of biologically
important reactions.

\begin{figure}[t]
\begin{centering}
\includegraphics[width=130mm]{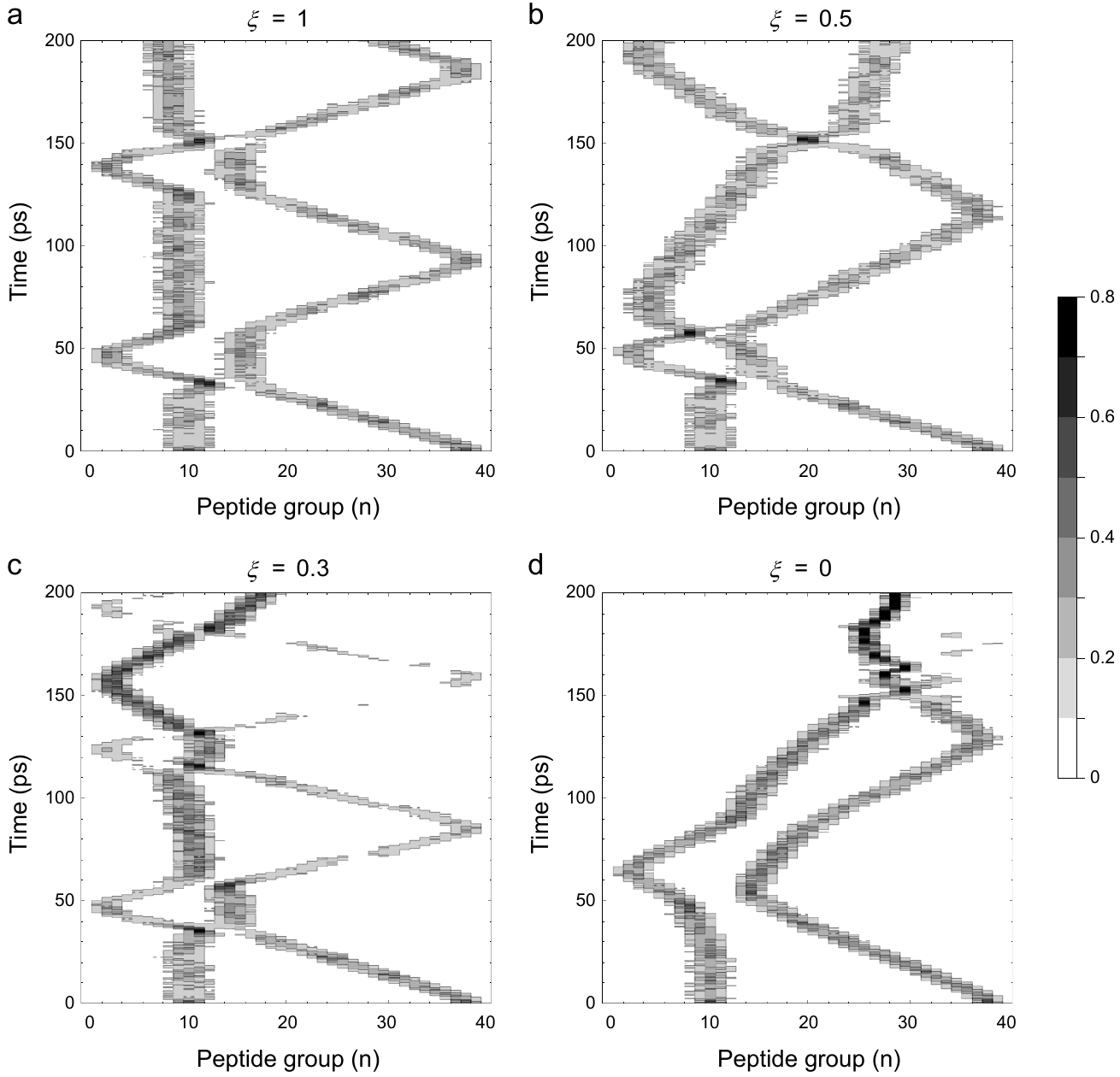}
\par\end{centering}

\caption{\label{fig:8}The quantum collision of one pinned and one moving Davydov
soliton with $Q=2$ quanta of amide~I energy visualized through the
expectation value of the exciton number operator $Q|a_{n}|^{2}$.
Decreasing the isotropy of exciton--phonon interaction leads to soliton
pinning after the collision: $\xi=1$ (a), $\xi=0.5$ (b), $\xi=0.3$
(c) and $\xi=0$ (d).}
\end{figure}

Interestingly, increasing the anisotropy of exciton--phonon interaction
by lowering $\xi<1$, is capable of inducing destructive quantum interference
at sites of soliton collision, so that these solitons appear to bounce off each
other without even touching (Fig.~\ref{fig:7}b). For $\xi<0.2$,
the soliton collision may lead to pinning of the soliton, which may
wobble out of the collision site for $\xi=0.1$ (Fig.~\ref{fig:7}c)
or remain pinned at the collision site for $\xi=0$ (Fig.~\ref{fig:7}d).
Thus, another potentially useful mechanism for persistent pinning
of energy at protein active centers may be the local modification
of exciton--phonon interaction anisotropy towards lower $\xi$ values.

Launching a propagating soliton from one end of the protein $\alpha$-helix
towards a pinned soliton in the interior of the helix, for $\xi=1$,
leads to destructive quantum interference of amide~I quantum probability
amplitudes $a_{n}$ at the collision site with the moving and pinned
soliton switching their roles between collisions (Fig.~\ref{fig:8}a).
In such collision scenario, lowering of the isotropy of exciton--phonon
interaction through the parameter $\xi$ leads to replacement of destructive
with constructive quantum interference at the collision sites (Figs.~\ref{fig:8}b,c).
For $\xi=0$, the collision of the two solitons at a site with constructive
quantum interference again leads to pinned soliton (Fig.~\ref{fig:8}d).
The irregular trajectories of solitons in the computer simulations
highlight the nonlinear dependence of the observed quantum dynamics
on the boundary conditions and point towards the emergence of quantum
chaos in regard to dynamics of quantum expectation values of amide
I energy.

\subsubsection{Three-soliton collisions}

\begin{figure}[t]
\begin{centering}
\includegraphics[width=130mm]{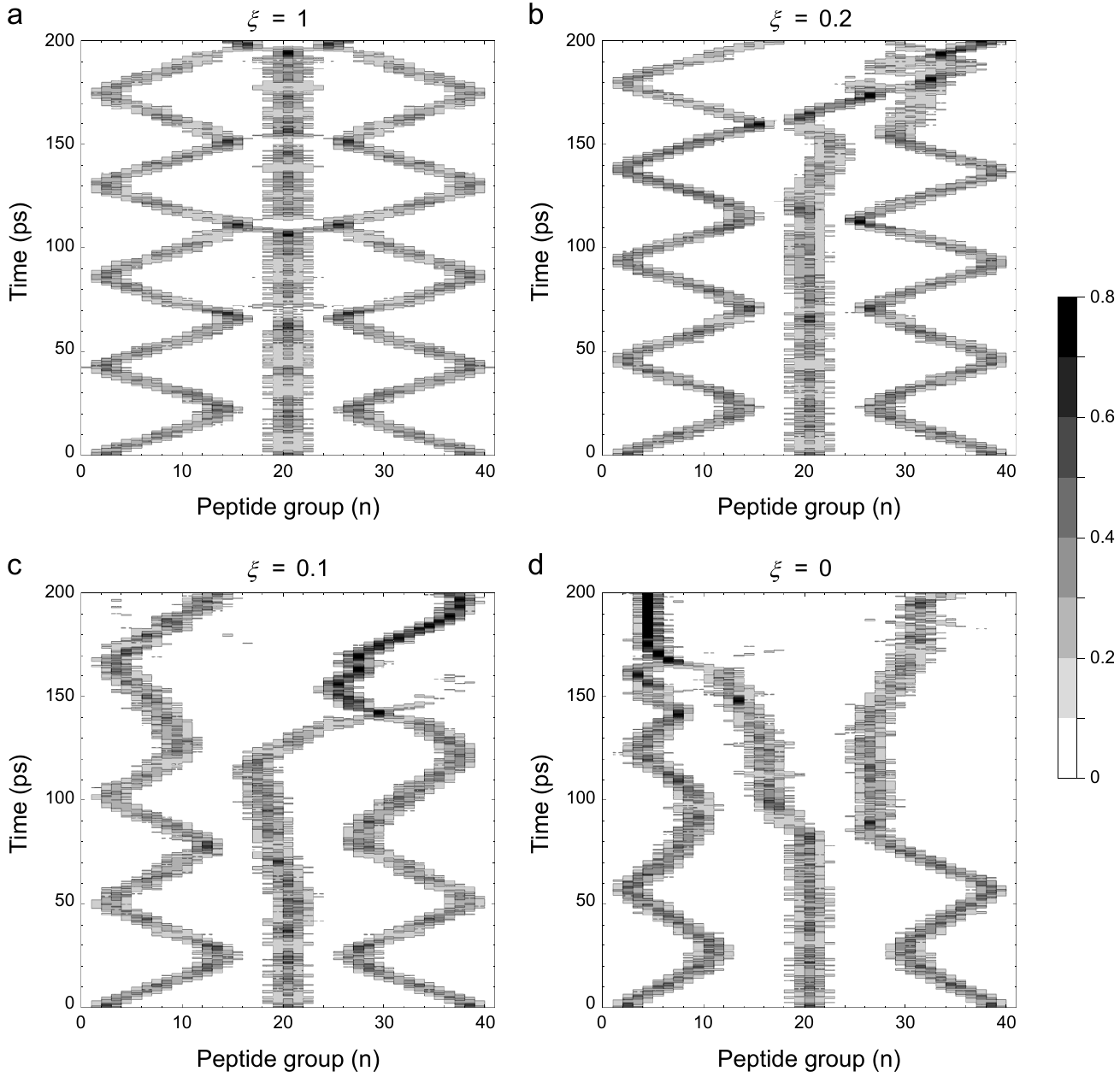}
\par\end{centering}

\caption{\label{fig:9}The quantum collision of one central pinned soliton and
two moving Davydov solitons with $Q=3$ quanta of amide~I energy visualized
through the expectation value of the exciton number operator $Q|a_{n}|^{2}$.
Destructive quantum interference prevents the solitons from colliding
and it appears that they repel each other. Decreasing the isotropy
of exciton--phonon interaction may lead to soliton pinning after the
collision: $\xi=1$ (a), $\xi=0.2$ (b), $\xi=0.1$ (c) and $\xi=0$
(d).}
\end{figure}

\begin{figure}[t]
\begin{centering}
\includegraphics[width=130mm]{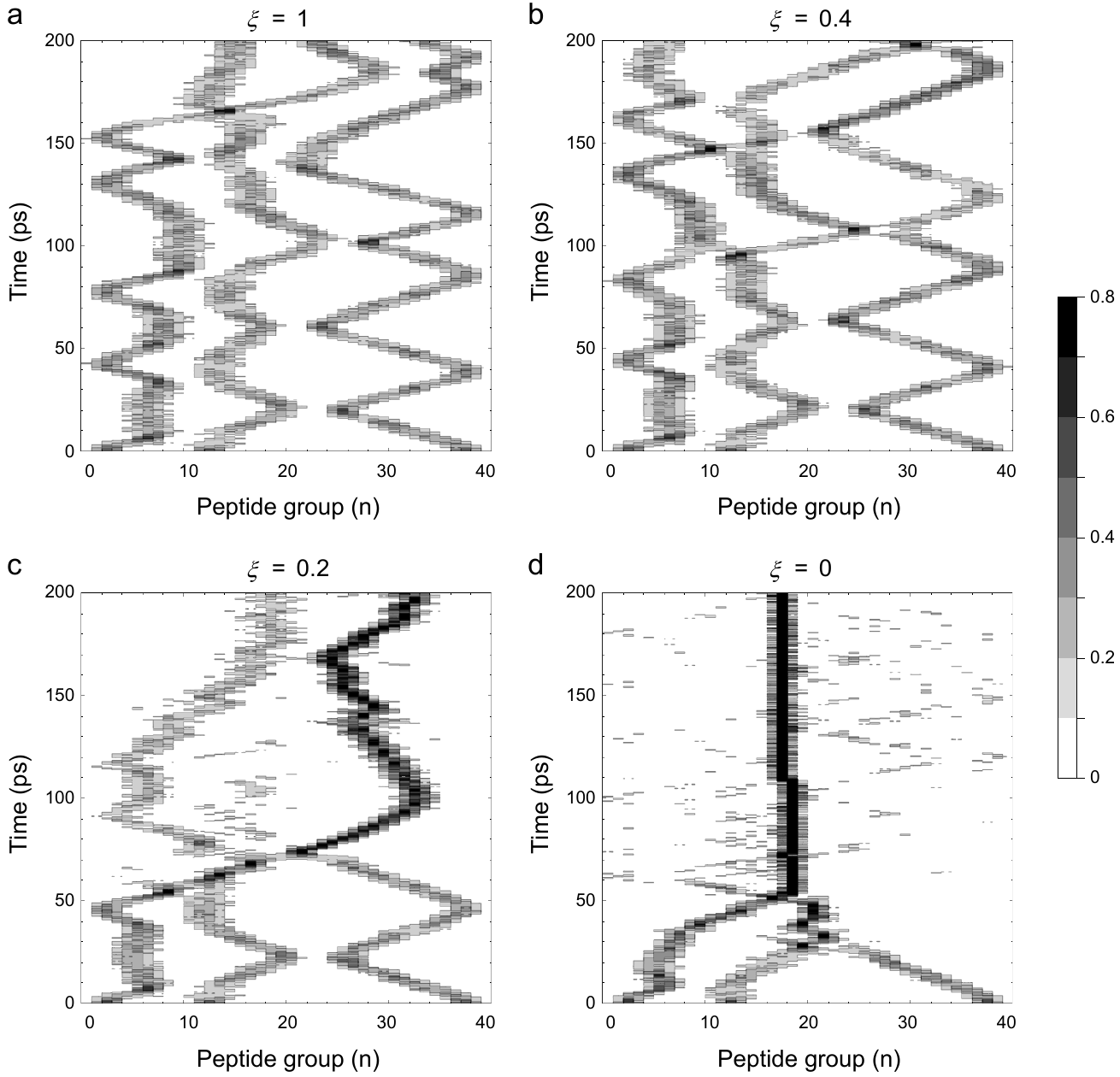}
\par\end{centering}

\caption{\label{fig:10} The quantum collision of one non-central pinned soliton
and two moving Davydov solitons with $Q=3$ quanta of amide~I energy
visualized through the expectation value of the exciton number operator
$Q|a_{n}|^{2}$. Destructive quantum interference prevents the solitons
from colliding and it appears that they repel each other. Decreasing
the isotropy of exciton--phonon interaction may lead to soliton pinning
after the collision: $\xi=1$ (a), $\xi=0.4$ (b), $\xi=0.2$ (c)
and $\xi=0$ (d).}
\end{figure}

Increasing the number of amide~I quanta to $Q=3$ allows for collision
of two moving solitons with one pinned soliton at a central position
(Fig.~\ref{fig:9}) or at a non-central position (Fig.~\ref{fig:10}).
In both of these cases, the pinned soliton appears to act as a divider,
which splits the protein $\alpha$-helix into compartments. The two
moving solitons, one launched from the N-end and the other launched
from the C-end, then reflect forth and back within the left compartment
or the right compartment, respectively (Fig.~\ref{fig:9}a).
Because the soliton width is spread over 5 peptide groups, which is an odd number, in the simulations with central soliton the protein length was set to $n_{\max}=41$ peptide groups in order to be able to perfectly center the soliton. Decreasing the isotropy of exciton--phonon interaction by lowering $\xi<1$ introduces asymmetry in the quantum dynamics with accidental drifts to the right (Fig.~\ref{fig:9}b) or to the left (Figs.~\ref{fig:9}c,d) of the central pinned soliton and renders irregular the trajectories of the two propagating solitons (Figs.~\ref{fig:9}b-d).
For $\xi\leq 0.1$, constructive quantum interference gives birth to a pinned soliton at the site of
collision (Fig.~\ref{fig:9}c,d) qualitatively reproducing the behavior
observed in two-soliton collisions (Figs.~\ref{fig:7}d and \ref{fig:8}d).

The three-soliton collision between two moving solitons and one non-central
pinned soliton exhibited even richer quantum dynamics (Fig.~\ref{fig:10}).
In such scenario, for $\xi=1$ the moving soliton launched from the
N-end of the protein $\alpha$-helix collided first with the pinned
soliton at $n=11-15$ switching the roles of the moving and pinned
soliton similarly to the two-soliton collision (cf. Fig.~\ref{fig:10}a vs Fig.~\ref{fig:8}a),
whereas the moving soliton launched from the C-end remained compartmentalized
and reflected forth and back without actual collision (cf. Fig.~\ref{fig:10}a
vs Fig.~\ref{fig:9}a). Again, decreasing the isotropy of exciton--phonon
interaction by lowering $\xi<1$ led to irregularities in the soliton
trajectories (Figs.~\ref{fig:10}c,d). For the completely anisotropic
exciton--phonon interaction $\xi=0$, the three solitons collided
in the center of the protein $\alpha$-helix producing a pinned
soliton that did not wobble around (Fig.~\ref{fig:10}d).

\begin{figure}[t]
\begin{centering}
\includegraphics[width=130mm]{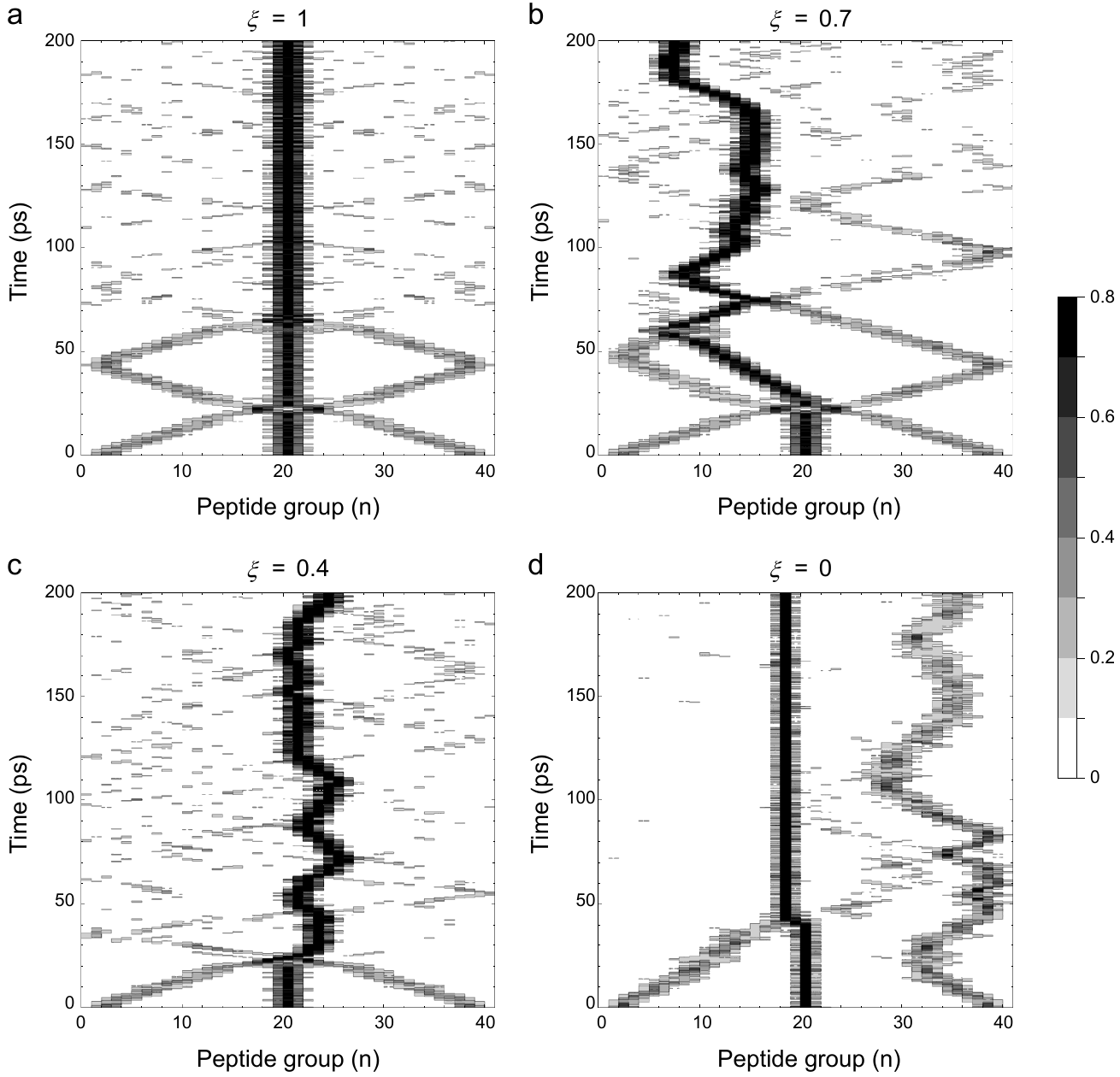}
\par\end{centering}

\caption{\label{fig:11} The quantum collision of one double central pinned soliton
and two moving Davydov solitons with $Q=4$ quanta of amide~I energy
visualized through the expectation value of the exciton number operator
$Q|a_{n}|^{2}$. Decreasing the isotropy of exciton--phonon interaction
leads to reduction in the shift induced by the collision and suppresses
the wobbling of the pinned soliton: $\xi=1$ (a), $\xi=0.7$ (b),
$\xi=0.4$ (c) and $\xi=0$ (d).}
\end{figure}

To further test the effect on compartmentalization of the protein
$\alpha$-helix by a central pinned soliton, we have doubled the central
soliton raising the total number of amide~I quanta to $Q=4$ (Fig.~\ref{fig:11}).
Interestingly, for $\xi=1$ the double central soliton stayed in the center where it devoured the single solitons feeding on their quantum probability amplitudes (Fig.~\ref{fig:11}a).
Decreasing the isotropy of exciton--phonon interaction by lowering $\xi<1$ introduced irregular
wobbling of the central soliton (Figs.~\ref{fig:11}b-d).
Thus, the collisions between single and double solitons in protein $\alpha$-helices
may have detrimental effects upon single solitons.

\subsection{\label{sub:4-4}Quantum tunneling of solitons}

\begin{figure}[t]
\begin{centering}
\includegraphics[width=130mm]{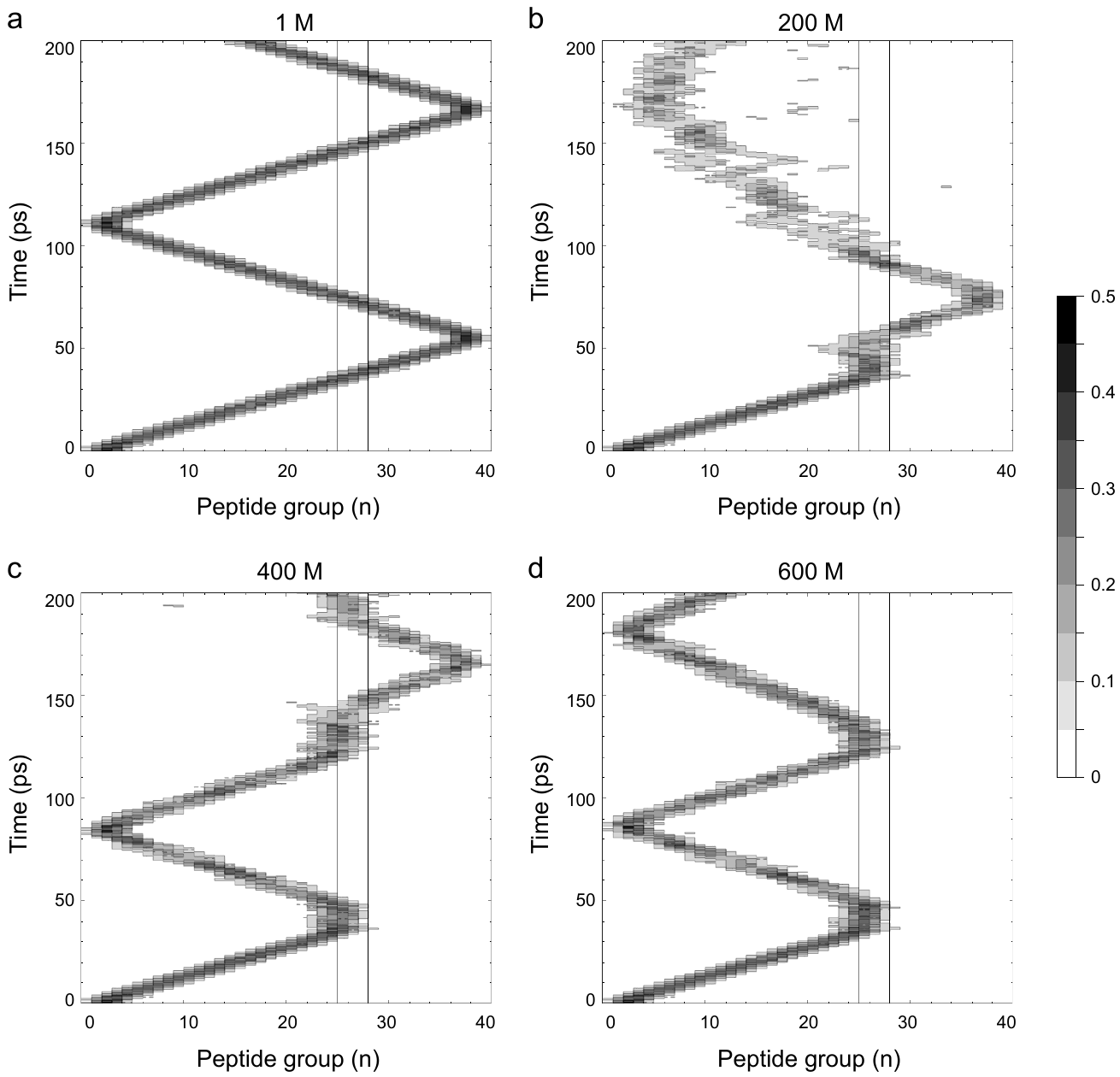}
\par\end{centering}

\caption{\label{fig:12} The quantum dynamics of a Davydov soliton with $Q=1$ quantum
of amide~I energy tunneling through or reflecting from a massive barrier
located over three peptide groups $n=26-28$ visualized through the
expectation value of the exciton number operator $Q|a_{n}|^{2}$ for
$\xi=1$. Increasing the mass of the barrier decreases the probability
of tunneling and increases the probability of reflection: no barrier
$1M$ (a), barrier in which each of the three peptide groups is with
effective mass of $200M$ (b), $400M$ (c) and $600M$ (d). The barrier
location is indicated with thin vertical lines.}
\end{figure}

Having verified the occurrence of quantum interference in soliton
collisions, we have turned our attention to the possibility of quantum
tunneling through massive barriers. Because the second of Davydov
equations \eqref{eq:gauge-2} depends on the mass of individual peptide
groups~$M_{n}$, in our previous work \cite{GeorgievGlazebrook2019b}
we have studied whether external protein clamps could act as massive
barriers by raising locally the effective mass of peptide groups inside
a protein $\alpha$-helix. In this latter case, we have shown that
single solitons with $Q=1$ that are wider behave as quasiparticles
with higher energy and are capable of tunneling through heavier barriers
in comparison with narrower solitons \cite{GeorgievGlazebrook2019b}.

\begin{figure}[t]
\begin{centering}
\includegraphics[width=130mm]{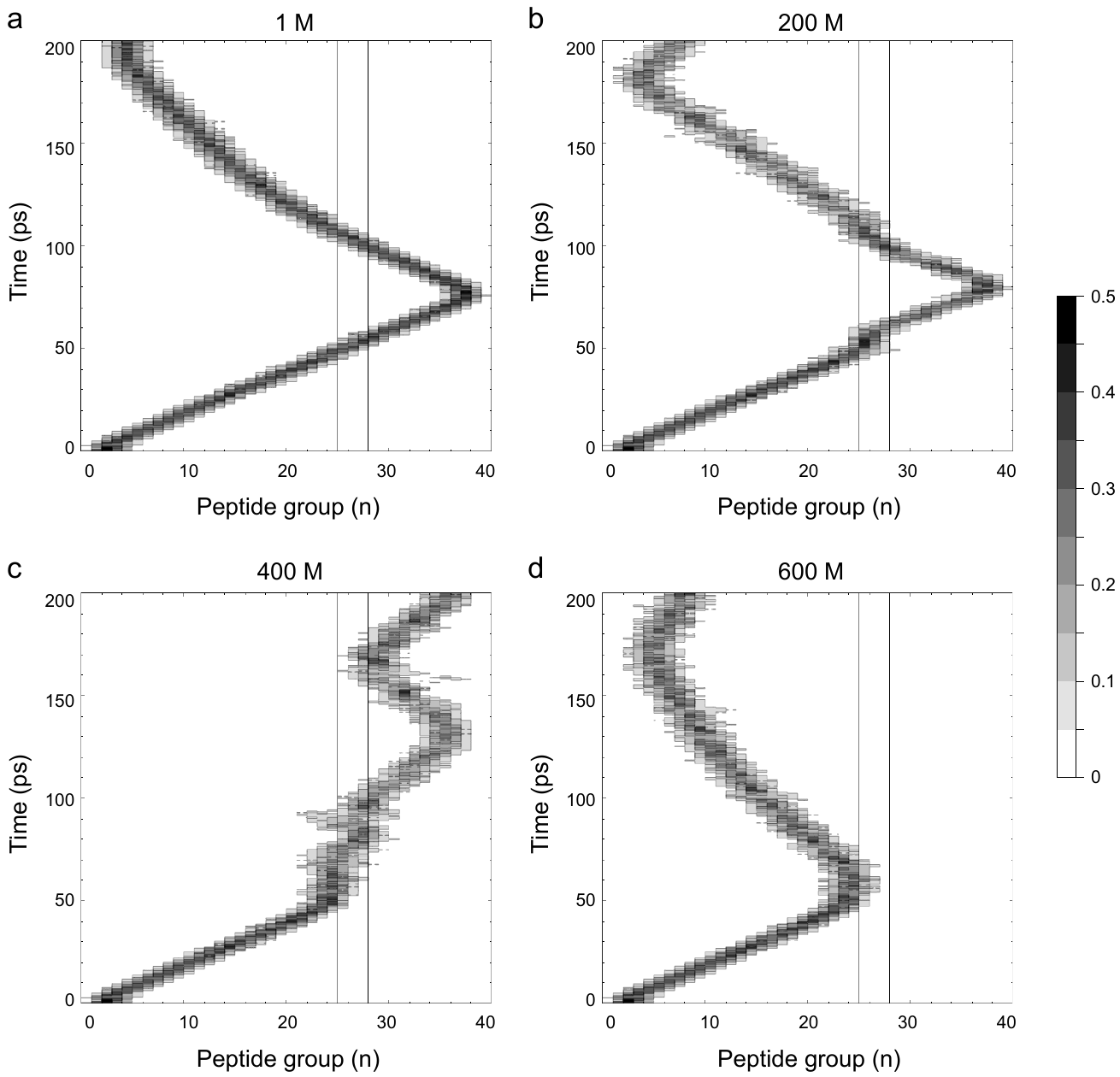}
\par\end{centering}

\caption{\label{fig:13} The quantum dynamics of a Davydov soliton with $Q=1$ quantum
of amide~I energy tunneling through or reflecting from a massive barrier
located over three peptide groups $n=26-28$ visualized through the
expectation value of the exciton number operator $Q|a_{n}|^{2}$ for
$\xi=0$. Increasing the mass of the barrier decreases the probability
of tunneling and increases the probability of reflection: no barrier
$1M$ (a), barrier in which each of the three peptide groups is with
effective mass of $200M$ (b), $400M$ (c) and $600M$ (d). The barrier
location is indicated with thin vertical lines.}
\end{figure}

Here, we have investigated the effect of increasing the number of
amide~I quanta for a fixed soliton width. To set a base for comparison,
we have first launched a single Davydov soliton with $Q=1$ by a Gaussian
pulse of amide~I energy distributed over 5 peptide groups at the N-end
of the protein $\alpha$-helix and aimed it at a massive barrier located
over three peptide groups $n=26-28$. This was repeated for the two
limiting cases, $\xi=1$ and $\xi=0$, of exciton--phonon interaction
isotropy (Figs.~\ref{fig:12} and \ref{fig:13}).

For $\xi=1$, the soliton readily tunneled through the barrier in
which each of the three peptide groups was with effective mass of
$200M$ (Fig.~\ref{fig:12}b), but got reflected from heavier barriers
with $400M$ (Fig.~\ref{fig:12}c) or $600M$ (Fig.~\ref{fig:12}d).
Thus, as it may be expected, increasing the mass of the barrier acts
analogously to increasing the height of a potential barrier thereby
reducing the probability of quantum tunneling of the soliton and increasing
the probability of its reflection.

\begin{figure}[t]
\begin{centering}
\includegraphics[width=130mm]{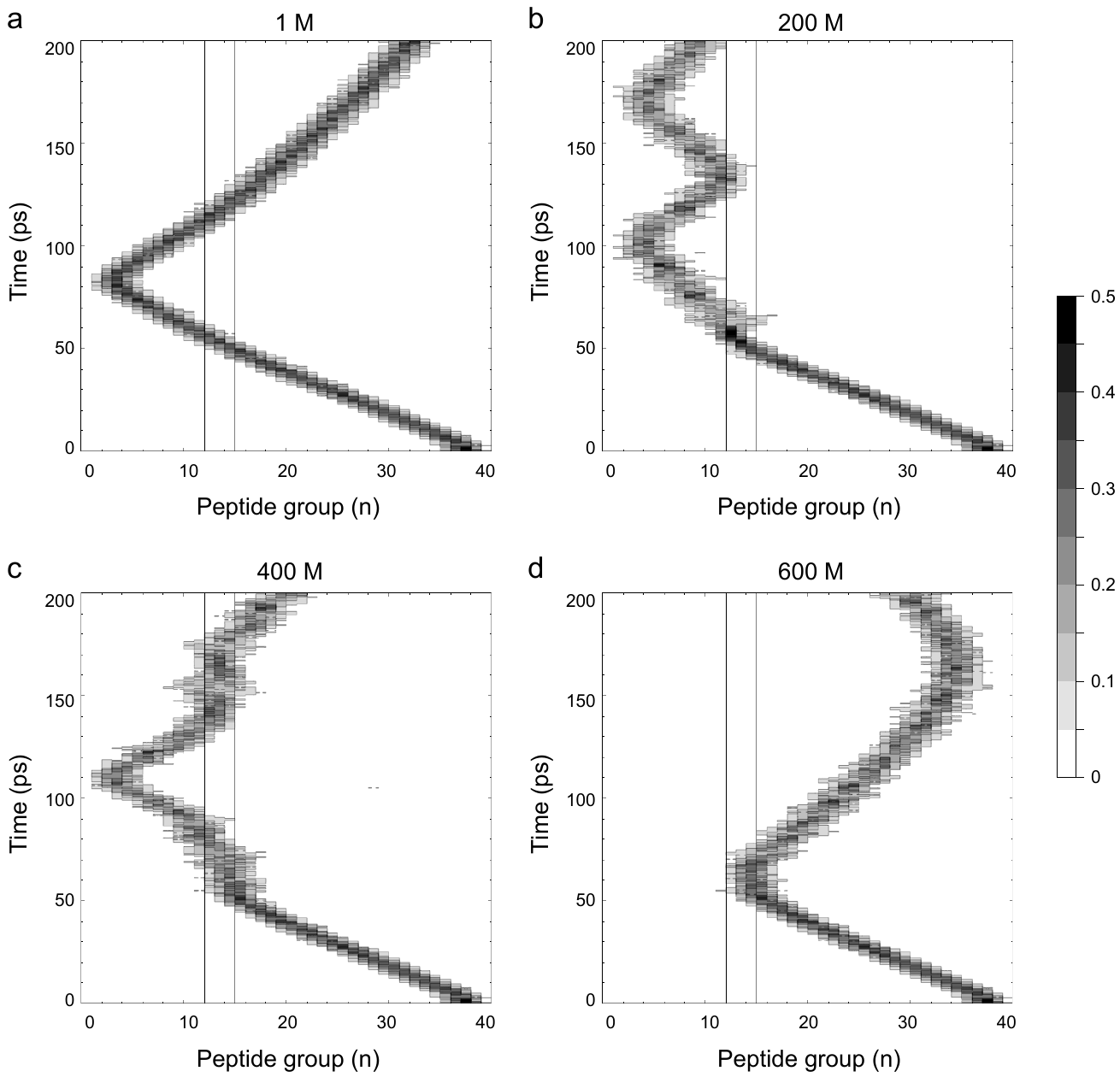}
\par\end{centering}

\caption{\label{fig:14} The quantum dynamics of a Davydov soliton with $Q=1$ quantum
of amide~I energy launched from the C-end of the $\alpha$-helix tunneling through or reflecting from a massive barrier
located over three peptide groups $n=13-15$ visualized through the
expectation value of the exciton number operator $Q|a_{n}|^{2}$ for
$\xi=0$. Increasing the mass of the barrier decreases the probability
of tunneling and increases the probability of reflection: no barrier
$1M$ (a), barrier in which each of the three peptide groups is with
effective mass of $200M$ (b), $400M$ (c) and $600M$ (d). The barrier
location is indicated with thin vertical lines.}
\end{figure}

For $\xi=0$, the soliton dynamics is not mirror symmetric with respect to launching from the N-end or the C-end of the protein $\alpha$-helix. Despite the lack of complete mirror symmetry, qualitatively the soliton behavior was similar: it was able to tunnel through both $200M$ barrier (Figs.~\ref{fig:13}b and \ref{fig:14}b) and $400M$ barrier (Figs.~\ref{fig:13}c and \ref{fig:14}c), but reflected from the $600M$ barrier (Figs.~\ref{fig:13}d and \ref{fig:14}d).
The soliton tunneling time through $200M$ barrier was also
faster for $\xi=0$, $22.4$ ps when launched from the N-end (Fig.~\ref{fig:13}b) and $29.8$ ps when launched from the C-end (Fig.~\ref{fig:14}b), compared with $33.2$ ps for $\xi=1$ (Fig.~\ref{fig:12}b), which is the same for launching from either end of the $\alpha$-helix.
Thus, consistently with our previously reported results for $Q=1$ \cite{GeorgievGlazebrook2019b} decreasing
the isotropy of exciton--phonon interaction by lowering $\xi$, decreases
the probability of soliton reflection from the barrier, increases
the probability of quantum tunneling of the soliton through the barrier
and reduces the tunneling time in the event of successful tunneling.

It should be noted that the soliton width does not appear to be significantly affected by the anisotropy of exciton--phonon interaction with $\xi<1$. The soliton width, defined by the spread of the exciton quantum probability amplitudes, is directly related to the expectation value of the exciton energy operator
\begin{equation}
\langle\Psi|\hat{H}_{\textrm{ex}}|\Psi\rangle=Q\sum_{n}\left[E_{0}|a_{n}|^{2}-J_{n+1}a_{n}^{*}a_{n+1}-J_{n}a_{n}^{*}a_{n-1}\right]
\end{equation}
The gauge transformation $a_{n}\to\bar{a}_{n}e^{-\frac{\imath}{\hbar}\int\gamma(t)dt}$ used to remove the highly oscillatory phase in Davydov's equations effectively sets $E_{0}=0$.
For the case when all $J_{n}$ are equal, the expectation of the exciton energy operator becomes
\begin{equation}
\langle\Psi|\hat{H}_{\textrm{ex}}|\Psi\rangle=-2QJ\sum_{n}\left[\textrm{Re}\left(a_{n}\right)\textrm{Re}\left(a_{n+1}\right)+\textrm{Im}\left(a_{n}\right)\textrm{Im}\left(a_{n+1}\right)\right]
\end{equation}
Thus, the soliton width is positively related to the absolute value of the
exciton expectation energy. When the soliton is narrowly focused onto
a single peptide group, then the expectation value is minimal $\left|\langle\Psi|\hat{H}_{\textrm{ex}}|\Psi\rangle\right|=0$,
whereas when the soliton is evenly spread over all peptide groups in the $\alpha$-helix spine
the expectation value is maximal $\left|\langle\Psi|\hat{H}_{\textrm{ex}}|\Psi\rangle\right|=2QJ$.

To correctly determine the soliton width in the discrete lattice, it
is necessary to consider the fact that, for most of the time, the soliton
is in the process of transition between neighboring peptide groups.
Thus, one would need a methodological rule that identifies time points
when the soliton is best positioned over the peptide groups for measuring
its width. Furthermore, on top of the soliton-induced deformation
of the lattice of hydrogen bonds there are superposed small disturbances
due to the phonon oscillations of the lattice, which may introduce
some noise on the exciton envelope of the soliton. Therefore, to find
out whether the soliton width changes in the course of the whole simulated
time period of 200 ps, we have divided the $\alpha$-helix spine into
35 overlapping local stretches of excition expectation energy given
by 5 terms in the sum for $\left|\langle\Psi|\hat{H}_{\textrm{ex}}|\Psi\rangle\right|$
as follows
\begin{equation}
\mathcal{E}_{i}\left(t\right)=2QJ\sum_{n=i-2}^{i+2}\left[\textrm{Re}\left(a_{n}\right)\textrm{Re}\left(a_{n+1}\right)+\textrm{Im}\left(a_{n}\right)\textrm{Im}\left(a_{n+1}\right)\right]
\end{equation}
The choice of $i\in[3,\ldots,37]$ ensures that the local stretches do not
extend outside the $\alpha$-helix spine ends, whereas the sum over
5 terms ensures that exciton probability amplitudes over 6 peptide
groups are captured. Then, the motion of the soliton leads to sequential
peaks of neighboring $\mathcal{E}_{i}\left(t\right)$ separated by
the time period needed for the soliton to travel from a position perfectly
centered on $\mathcal{E}_{i}\left(t\right)$, to a spatially translated
position when the soliton is centered on $\mathcal{E}_{i+1}\left(t\right)$.
Variation of the time intervals between peaks of neighboring local
stretches $\mathcal{E}_{i}\left(t\right)$ indicates a variation in
the soliton speed, whereas a non-zero slope of the trend line of exciton
energy peaks will be an indication of changing soliton width. For
example, a positive slope of the trend line of exciton energy peaks
will indicate that the soliton becomes wider, whereas a negative trend line
of exciton energy peaks will indicate that the soliton becomes narrower.
This will hold true as long as the soliton width fits inside the stretches
$\mathcal{E}_{i}\left(t\right)$ (hence no contributions to the exciton
expectation energy will be trimmed) and explains why we have set the
length of the streches to be 6 peptide groups given that the initial
soliton width is 5 peptide groups.

\begin{figure}[t!]
\begin{centering}
\includegraphics[width=130mm]{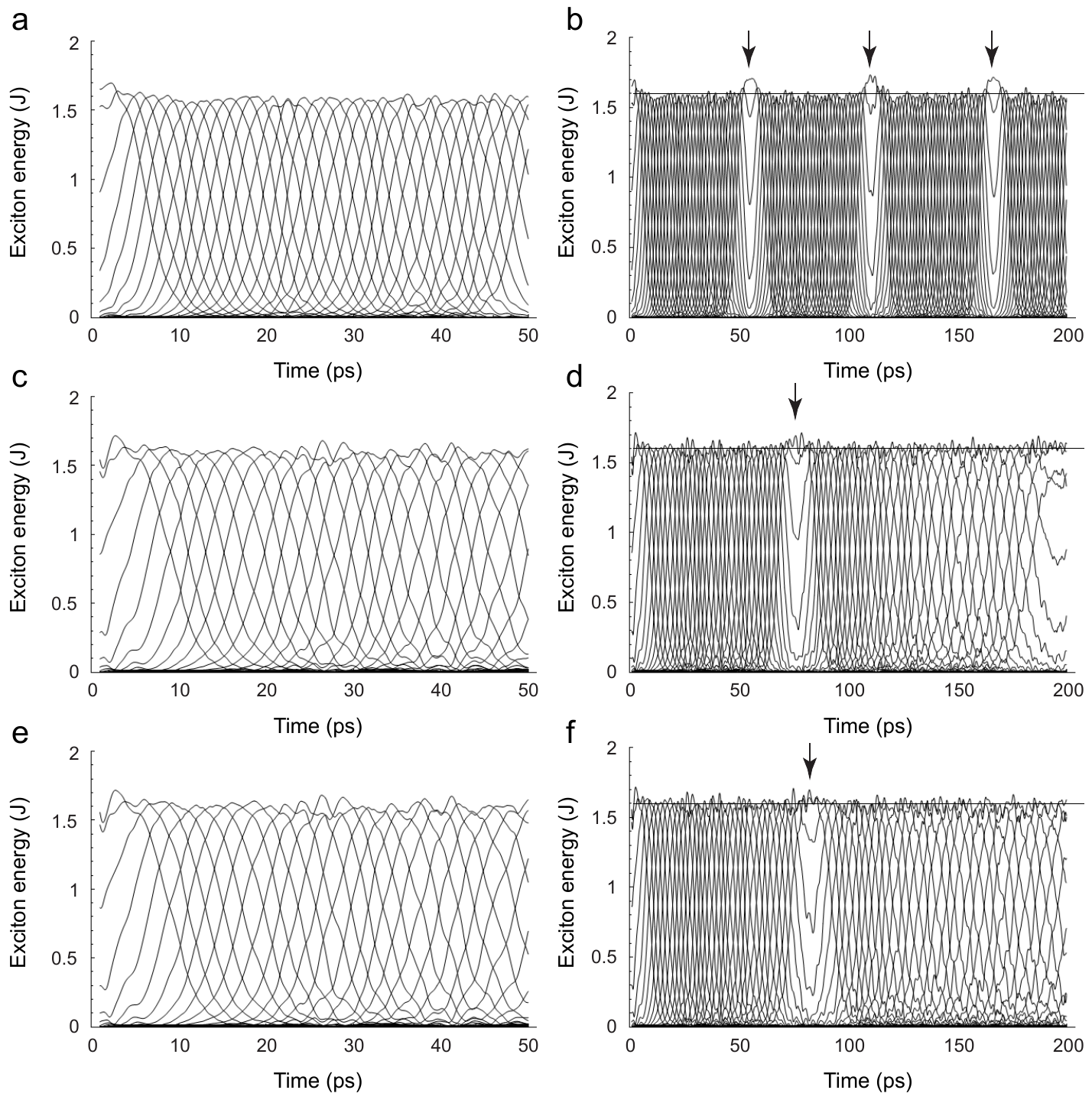}
\par\end{centering}

\caption{\label{fig:15} The quantum dynamics of the local exciton energy (measured in units of $J = 0.155$~zJ) computed for stretches of 5 terms $\mathcal{E}_{i}\left(t\right)$ in the sum for $|\langle\Psi|\hat{H}_\textrm{ex}|\Psi\rangle|$ of a Davydov soliton with $Q=1$ quantum of amide~I energy for different values of exciton--phonon interaction isotropy $\xi$.
(a-b)~Exciton energy for $\xi=1$ soliton launched from the N-end of the $\alpha$-helix.
(c-d)~Exciton energy for $\xi=0$ soliton launched from the N-end of the $\alpha$-helix.
(e-f)~Exciton energy for $\xi=0$ soliton launched from the C-end of the $\alpha$-helix.
The horizontal trend lines for $\mathcal{E}_{i}\left(t\right)$ peaks at~$1.6 J$ show that the changes in soliton velocity are not accompanied by spread in the envelope of exciton quantum probability amplitudes. Arrows indicate soliton reflection from the $\alpha$-helix ends.}
\end{figure}

To assess the effects of $\xi$ on soliton width and velocity, we have compared the simulations with $Q=1$ reported in Figs.~\ref{fig:12}a, \ref{fig:13}a and \ref{fig:14}a for $\xi=1$ launched from the N-end, and for $\xi=0$ launched from the N-end or C-end, respectively. The distances between consecutive $\mathcal{E}_{i}\left(t\right)$ peaks are shorter for $\xi=1$ compared with $\xi=0$ (Figs.~\ref{fig:15}a,c,e) consistent with higher velocities of $\xi=1$ solitons. While the soliton velocity is relatively constant for $\xi=1$ for the whole simulation period of 200~ps (Fig.~\ref{fig:15}b), it appears to slow down for $t>120$~ps for $\xi=0$ (Figs.~\ref{fig:15}d,f). This retardation of the $\xi=0$ solitons, however, is not accompanied by any significant spread in the envelope of exciton quantum probability amplitudes, since the trend lines for $\mathcal{E}_{i}\left(t\right)$ peaks remain horizontal at~$1.6 J$ (Figs.~\ref{fig:15}d,f). Thus, the mechanism behind the varying soliton velocity for $\xi=0$ could be quantum interference within the finite length of the discrete lattice. Also, the ambient noise on the exciton envelope, resulting from the phonon lattice oscillations, is greater for $\xi=0$, and is probably due to manifestly nonlinear effects dependent on $\chi_r$, because to achieve the average $\bar{\chi}=35$~pN, the right exciton--phonon coupling becomes $\chi_r=70$~pN given that $\chi_l=0$~pN. In contrast, for $\xi=1$  the average $\bar{\chi}=35$~pN is obtained with $\chi_r=35$~pN and $\chi_l=35$~pN.

\begin{figure}[t]
\begin{centering}
\includegraphics[width=130mm]{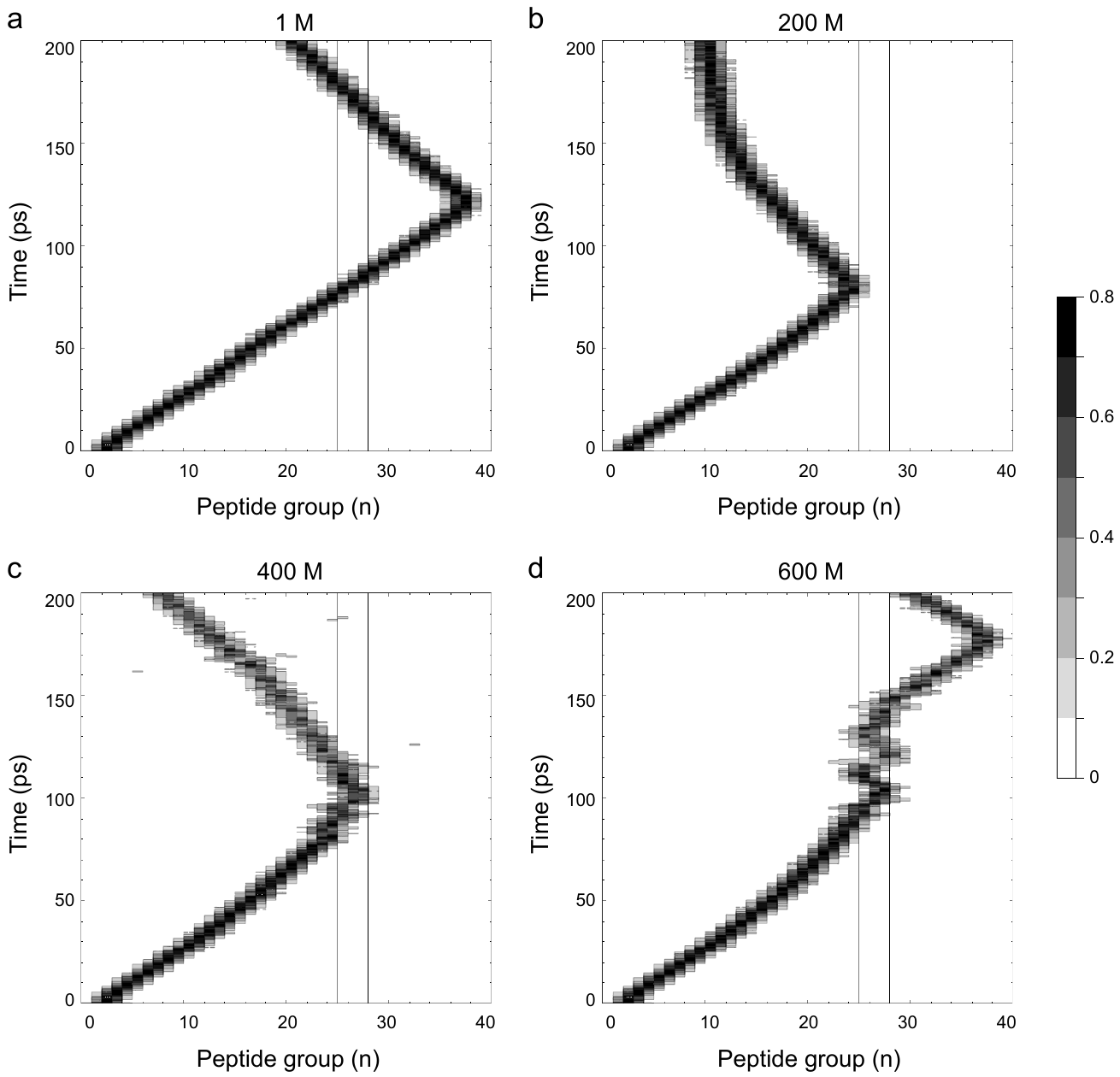}
\par\end{centering}

\caption{\label{fig:16} The quantum dynamics of a Davydov soliton with $Q=2$ quanta
of amide~I energy tunneling through or reflecting from a massive barrier
located over three peptide groups $n=26-28$ visualized through the
expectation value of the exciton number operator $Q|a_{n}|^{2}$ for
$\xi=1$. Increasing the mass of the barrier may increase the interaction
time with the barrier and increase the probability of tunneling through
the barrier: no barrier $1M$ (a), barrier in which each of the three
peptide groups is with effective mass of $200M$ (b), $400M$ (c)
and $600M$ (d). The barrier location is indicated with thin vertical
lines.}
\end{figure}

Doubling the launched soliton by increasing the amide~I quanta to
$Q=2$, for $\xi=1$, reveals a soliton reflection from $200M$ barrier
(Fig. \ref{fig:16}b) or $400M$ barrier (Fig. \ref{fig:16}c), but
remarkable tunneling phenomena through the much more massive $600M$
barrier (Fig. \ref{fig:16}d). This quantum behavior provides an indication
that the tunneling of the Davydov soliton through the massive barrier
could be analogous to a massive quantum particle tunneling through a
potential barrier whose potential barrier height $V_{0}$ is lower
than the energy $E_{0}$ of the particle \cite{Landau1965,GeorgievGlazebrook2018}.

\begin{figure}[t]
\begin{centering}
\includegraphics[width=130mm]{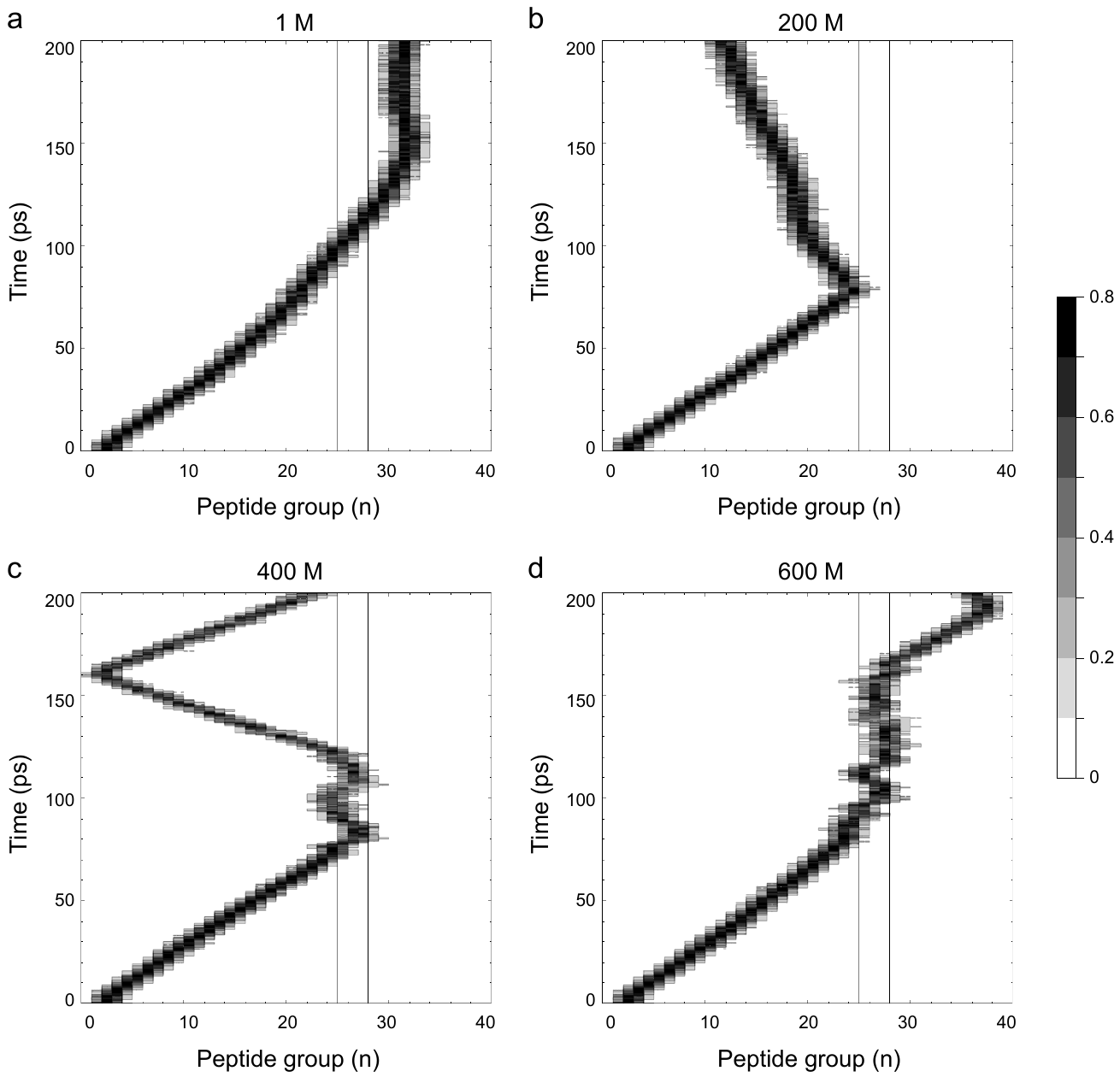}
\par\end{centering}

\caption{\label{fig:17} The quantum dynamics of a Davydov soliton with $Q=2$ quanta
of amide~I energy tunneling through or reflecting from a massive barrier
located over three peptide groups $n=26-28$ visualized through the
expectation value of the exciton number operator $Q|a_{n}|^{2}$ for
$\xi=0.7$. Increasing the mass of the barrier may increase the interaction
time with the barrier and increase the probability of tunneling through
the barrier: no barrier $1M$ (a), barrier in which each of the three
peptide groups is with effective mass of $200M$ (b), $400M$ (c)
and $600M$ (d). The barrier location is indicated with thin vertical
lines.}
\end{figure}

For a massive quantum particle with mass $m$ and energy $E_{0}$
tunneling through rectangular potential barrier
\begin{equation}
V(x)=V_{0}\left[\Theta\left(x-x_{1}\right)-\Theta\left(x-x_{2}\right)\right]
\end{equation}
with height $V_{0}$ and width $\Delta x=x_{2}-x_{1}$, where $\Theta\left(x\right)=\frac{d}{dx}\max\left\{ x,0\right\} $
is the Heaviside step function, the transmission coefficient $T$
is determined from different quantum mechanical expressions depending
on the magnitude of $V_{0}$ with respect to $E_{0}$, which is required
to avoid the appearance of imaginary wavenumber $k_{2}$ as follows.

Case I: If $V_{0}>E_{0}$, on setting $\hbar k_{1}=\sqrt{2mE_{0}}$
and $\hbar k_{2}=\sqrt{2m\left(V_{0}-E_{0}\right)}$ \cite[p. 75--80]{Landau1965},
the analytic derivation of the transmission coefficient gives
\begin{equation}
T=\frac{4k_{1}^{2}k_{2}^{2}}{\left(k_{1}^{2}+k_{2}^{2}\right)^{2}\sinh^{2}\left(\Delta xk_{2}\right)+4k_{1}^{2}k_{2}^{2}}\label{eq:case-1}
\end{equation}

Case II: If $V_{0}<E_{0}$, on setting $\hbar k_{1}=\sqrt{2mE_{0}}$
and $\hbar k_{2}=\sqrt{2m\left(E_{0}-V_{0}\right)}$ \cite[p. 75--80]{Landau1965},
the analytic derivation of the transmission coefficient gives
\begin{equation}
T=\frac{4k_{1}^{2}k_{2}^{2}}{\left(k_{1}^{2}-k_{2}^{2}\right)^{2}\sin^{2}\left(\Delta xk_{2}\right)+4k_{1}^{2}k_{2}^{2}}\label{eq:case-2}
\end{equation}

In the latter case, for $\Delta x>\frac{1}{k_{2}}$, it is indeed possible
to observe larger transmission coefficient $T$ for larger $V_{0}$
due to the occurrence of quantum interference effects. Thus, the result of the simulation
reported in Fig.~\ref{fig:16}, supports the conclusion that
doubling the amide~I quanta in the double soliton with $Q=2$, is analogous
to raising the energy $E_{0}$ of massive particle that now faces
a potential barrier with height $V_{0}<E_{0}$.

Because a particular effect of decreasing the isotropy of exciton--phonon
interaction by lowering $\xi$, is to pin down the soliton, for the double
soliton with $Q=2$, we were unable to test quantum tunneling through
the massive barrier for $\xi<0.7$. In the case of $\xi=0.7$, the
double soliton reflected from $200M$ barrier (Fig. \ref{fig:17}b)
or $400M$ barrier (Fig. \ref{fig:17}c), yet it tunneled through
the much more massive $600M$ barrier (Fig. \ref{fig:17}d). This
behavior was qualitatively similar to the completely isotropic case
with $\xi=1$. However, in comparison with the case $\xi=1$, for which
the tunneling time was $71.1$ ps (Fig. \ref{fig:16}d), the presence
of some anisotropy for $\xi=0.7$ delayed the passage through the
barrier with a tunneling time of $92.9$ ps (Fig. \ref{fig:17}d).
Thus, the faster tunneling time for higher $\xi$ in double solitons
with $Q=2$ differs from the observed tunneling times for single solitons
with $Q=1$, and highlights the non-classical nature of soliton transmission
through a potential barrier with $V_{0}<E_{0}$ due to manifested quantum
interference effects in the transmission coefficient \eqref{eq:case-2}.

\subsection{\label{sub:4-5}Disorder effects on soliton stability}

\begin{figure}[t]
\begin{centering}
\includegraphics[width=130mm]{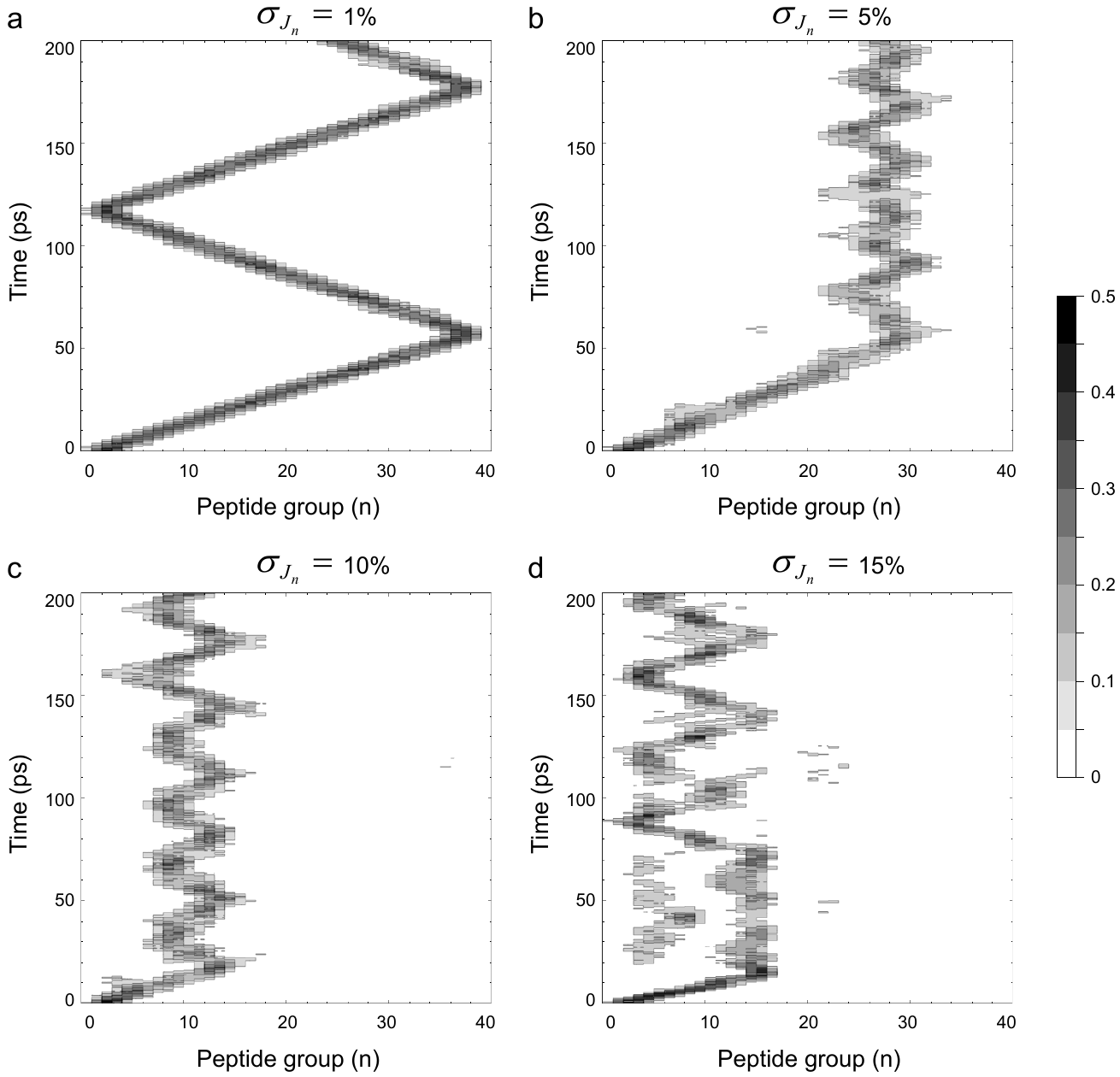}
\par\end{centering}

\caption{\label{fig:18} The quantum dynamics of a Davydov soliton with $Q=1$ quantum
of amide~I energy for protein $\alpha$-helix with randomly variable
$J_{n}$ dipole--dipole coupling energy between amide~I oscillators
visualized through the expectation value of the exciton number operator
$Q|a_{n}|^{2}$ for $\xi=1$. Non-uniformity of $J_{n}$ was quantified
by the standard deviation $\sigma_{J_{n}}$ expressed as percentage
from the mean value $J$: $\sigma_{J_{n}}=1\%$ (a), $\sigma_{J_{n}}=5\%$
(b), $\sigma_{J_{n}}=10\%$ (c) and $\sigma_{J_{n}}=15\%$ (d). Increasing
the non-uniformity of $J_{n}$ destabilizes the soliton and traps it to
a region flanked by low $J_{n}$ values.}
\end{figure}

The creation of soliton solutions by the system of Davydov equations
\eqref{eq:gauge-1} and \eqref{eq:gauge-2} is made possible by the
highly ordered protein $\alpha$-helix structure, which is reflected
in the construction of the Hamiltonian \eqref{eq:Hamiltonian}. Because
both living and non-living quantum physical systems are subject to
the same fundamental quantum physical laws, we aimed at elucidating
the importance of biological order \cite{Schrodinger1977,Jordan1941,Beyler1996}
for outlining the quantum boundaries of life.

In most quantum chemistry applications, the Born--Oppenheimer approximation
\cite{Born1927} allows the Schrödinger equation for a biomolecule
to be separated it into two equations: i) an electronic Schrödinger equation,
and ii) a nuclear Schrödinger equation. Only the positions of the much
heavier atomic nuclei (not their momenta) generate the potential that
enters in the Hamiltonian for solving the Schrödinger equation for
the electrons \cite{Lowe2005}. Once the solution $\psi_{e}$ of the
electronic Schrödinger equation is found, it is used to provide the
potential energy function for the nuclear motion. The obtained vibrational
nuclear wavefunctions~$\psi_{n}$ with corresponding energies solve the nuclear
Schrödinger equation. The total wavefunction $\Psi=\psi_{e}\psi_{n}$
for the biomolecule is then composed as a product of the electronic
wavefunction~$\psi_{e}$ and the nuclear wave function~$\psi_{n}$.
Further simplification of quantum mechanical calculations could be achieved with the Crude Born--Oppenheimer Approximation where the equilibrium separation of the nuclei is employed at all times.

\begin{figure}[t]
\begin{centering}
\includegraphics[width=130mm]{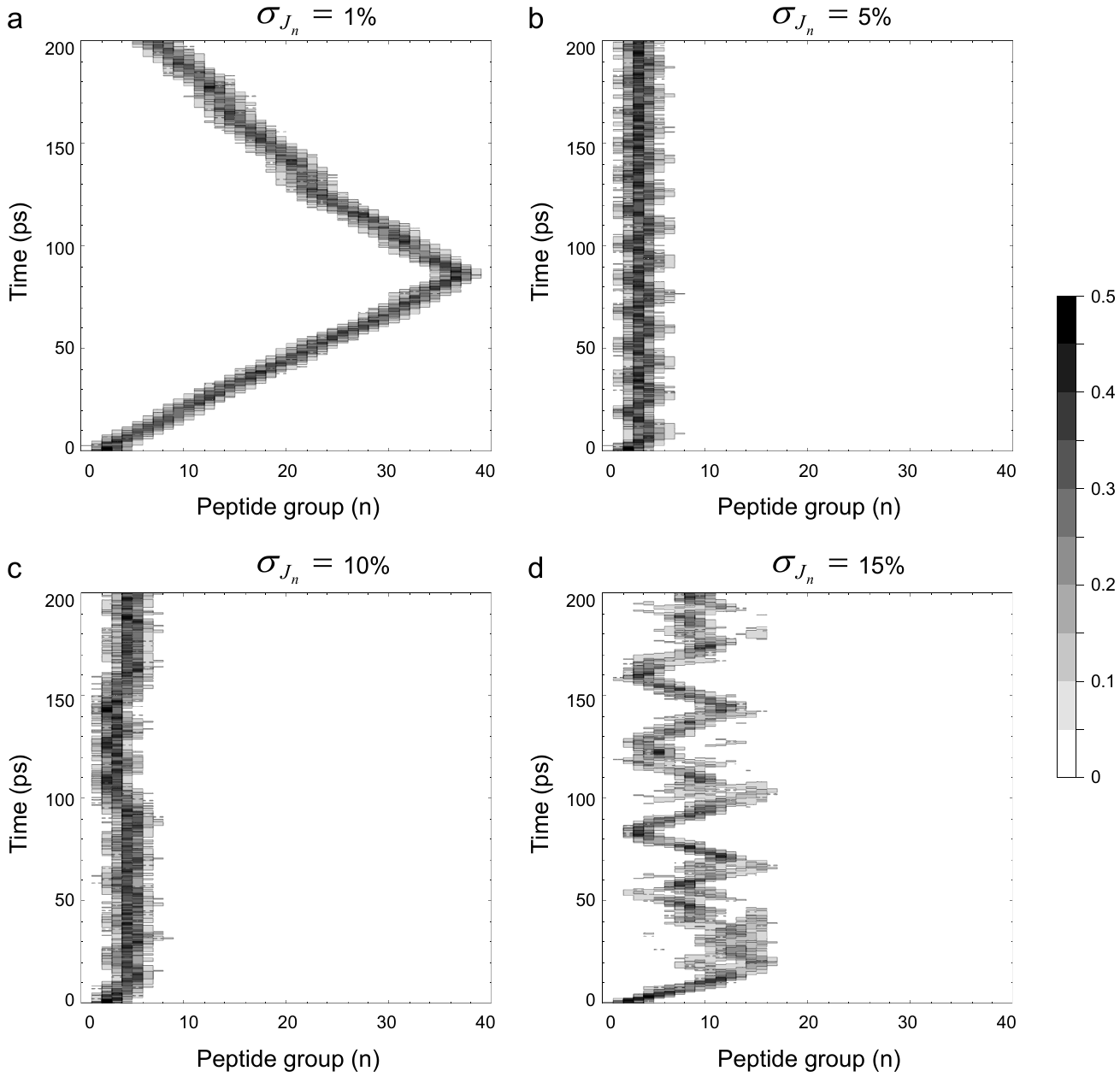}
\par\end{centering}

\caption{\label{fig:19} The quantum dynamics of a Davydov soliton with $Q=1$ quantum
of amide~I energy for protein $\alpha$-helix with randomly variable
$J_{n}$ dipole--dipole coupling energy between amide~I oscillators
visualized through the expectation value of the exciton number operator
$Q|a_{n}|^{2}$ for $\xi=0$. Non-uniformity of $J_{n}$ was quantified
by the standard deviation $\sigma_{J_{n}}$ expressed as percentage
from the mean value $J$: $\sigma_{J_{n}}=1\%$ (a), $\sigma_{J_{n}}=5\%$
(b), $\sigma_{J_{n}}=10\%$ (c) and $\sigma_{J_{n}}=15\%$ (d). Increasing
the non-uniformity of $J_{n}$ destabilizes the soliton.}
\end{figure}

The upshot of the Crude Born--Oppenheimer approximation manifests
in the original Davydov Hamiltonian \eqref{eq:Hamiltonian} in the guise of uniformity
of dipole--dipole coupling energies $J_{n}$ between neighboring amide
I oscillators, which are all set to $J=0.155$ zJ \cite{Davydov1976,Davydov1979,Davydov1982,Davydov1986}.
This introduces a high degree of stability in the biological system. In
order to study the effects of instability on Davydov solitons, we have
drawn randomly the value of each $J_{n}$ from a Gaussian distribution
with mean value $J$, and standard deviation $\sigma_{J_{n}}$, defined as
a percentage of~$J$. To avoid reporting an outlier quantum dynamics,
we have simulated at least 5 randomly drawn distributions for a given
$\sigma_{J_{n}}$ and presented the outcome of a simulation run, which
had been supported by another visually similar outcome.

\begin{figure}[t]
\begin{centering}
\includegraphics[width=130mm]{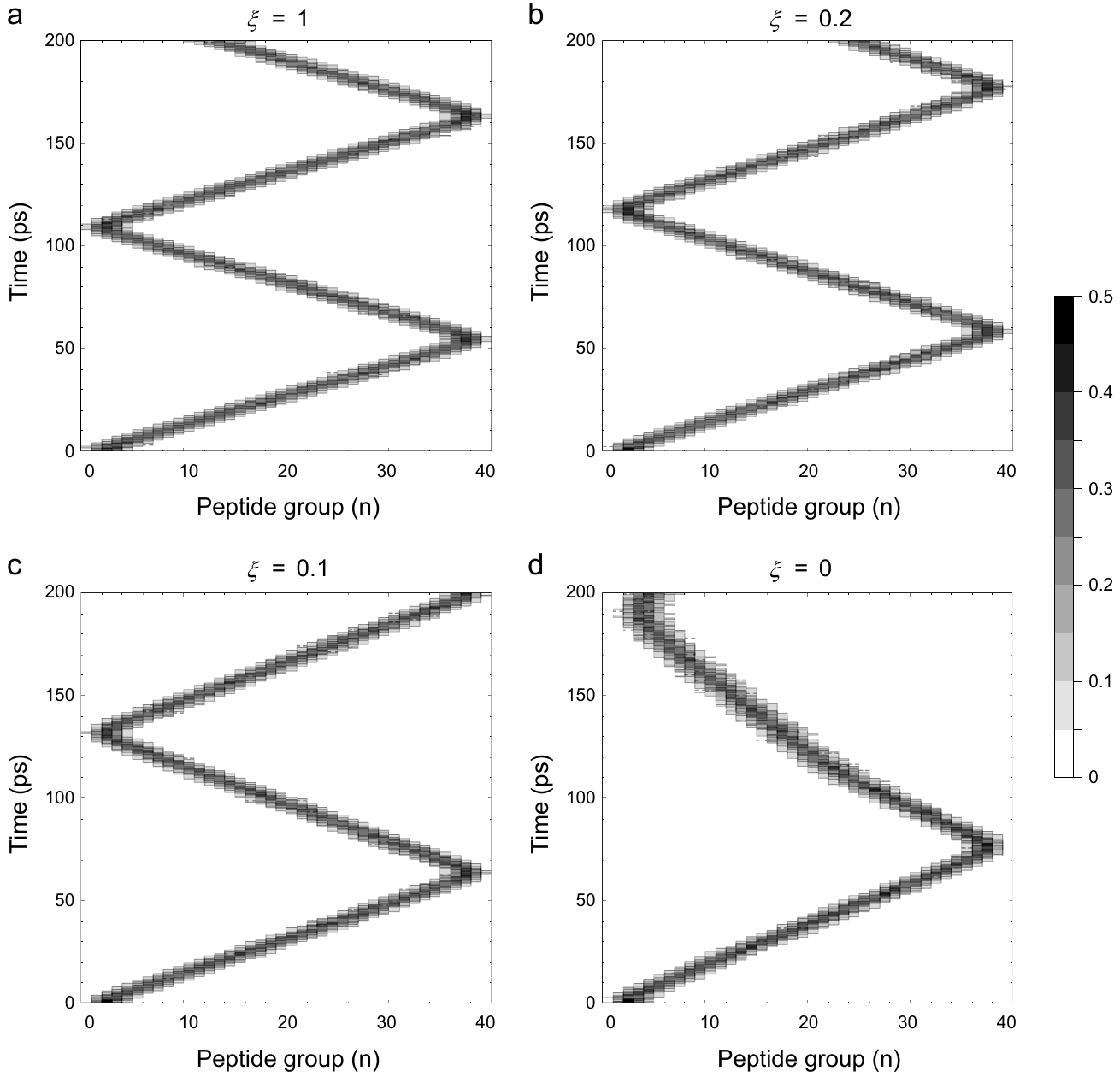}
\par\end{centering}

\caption{\label{fig:20}The quantum dynamics of a Davydov soliton with $Q=1$ quantum
of amide~I energy for protein $\alpha$-helix with nonuniform
$M_{n}$ peptide group masses (standard deviation $\sigma_{M_{n}}$
is 25\% from the mean value $M=1.9\times10^{-25}$~kg) visualized through the expectation
value of the exciton number operator $Q|a_{n}|^{2}$ for different
values of the isotropy of exciton--phonon interaction: $\xi=1$ (a),
$\xi=0.2$ (b), $\xi=0.1$ (c) and $\xi=0$ (d).}
\end{figure}

For a protein $\alpha$-helix with completely isotropic exciton--phonon
interaction $\xi=1$, the presence of a dipole--dipole coupling disorder
$\sigma_{J_{n}}<5\%$ was not detrimental for the moving soliton (Fig.~\ref{fig:18}a).
When the disorder was increased between $\sigma_{J_{n}}=5\%$ (Fig.~\ref{fig:18}b)
and $\sigma_{J_{n}}=10\%$ (Fig.~\ref{fig:18}c), the soliton propagated
along the protein $\alpha$-helix until it was trapped inside a region
flanked by low $J_{n}$ values. Visually, the quantum dynamics of
the soliton resembled compartmentalization inside a short protein
$\alpha$-helix rather than pinning. For greater disorder $\sigma_{J_{n}}=15\%$,
features of soliton disintegration were observed even though part
of the soliton still persisted (Fig.~\ref{fig:18}d).

For a protein $\alpha$-helix with completely anisotropic exciton--phonon
interaction $\xi=0$, the presence of dipole--dipole coupling disorder
had similar effects, however the soliton appeared to be pinned for
$\sigma_{J_{n}}=5\%$ (Fig.~\ref{fig:19}b) and $\sigma_{J_{n}}=10\%$
(Fig.~\ref{fig:19}c). The destabilization of the soliton at $\sigma_{J_{n}}=15\%$
was present (Fig.~\ref{fig:19}d), albeit a bit weaker compared with
the completely isotropic $\xi=1$ case. Thus, the order and uniformity
of dipole--dipole coupling energies between neighboring amide~I oscillators
is pivotal for the highly efficient, dissipationless, solitonic transport
of energy by protein $\alpha$-helices. In addition, cooperative effects
between backbone dipole--dipole interactions are instrumental for
the formation of secondary and supersecondary structures of proteins,
and after successful folding keep proteins in their folded structural
state \cite{Ganesan2014}, while noting that protein conformational transitions, as microscopic biochemical processes, {\it a fortiori}, involve quantum states (see e.g. \cite{Cruzeiro2011}).

To illustrate the fact that not all types of disorder are equally
adverse to soliton dynamics, we have also randomly varied the amino
acid masses $M_{n}$ within the maximal biochemical range of $\sigma_{M_{n}}=25\%$
in a protein $\alpha$-helix. For all values of $\xi$, the solitons
readily propagated along the $\alpha$-helix spine (Fig.~\ref{fig:20})
and were virtually indistinguishable from the case with uniform amino
acid masses (cf. Fig.~\ref{fig:12}a vs Fig.~\ref{fig:20}a and
Fig.~\ref{fig:13}a vs Fig.~\ref{fig:20}d).

\section{Conclusions}

The importance of proteins for maintaining life cannot be overstated
\cite{Alberts2013,Milner2019}. The majority of biological processes
are catalyzed or performed by proteins, and the main part of the coding
DNA in the genome is dedicated to storing hereditary information for
the production of proteins. The underlying mechanisms behind the versatility
of protein function, however, remained elusive within the deterministic
clockwork structures of classical physics \cite{Schrodinger1977,Jordan1941}.
In an attempt to address the perceived theoretical crisis in bioenergetics, Alexander Davydov, in ~1973, proposed that the amide~I quanta of peptide vibrational
energy (C=O stretching) might become self-localized through interactions
with lattice phonons in protein $\alpha$-helices \cite{Davydov1973}.
The original Davydov model was analytically studied with the use of
the continuum approximation or simulated numerically \cite{Brizhik1983,Brizhik1988,Brizhik1993,Brizhik1995,Brizhik2004,Brizhik2006,Brizhik2010,Davydov1976,Davydov1979,Davydov1982,Davydov1986,Davydov1987,Davydov1988},
and then further generalized to include explicit treatment of multiple
amide~I quanta and the possible anisotropies of various physical parameters
in realistic protein models \cite{Cruzeiro1988,Cruzeiro1994,Cruzeiro1997,Cruzeiro2009,Forner1990,Forner1991c,Kerr1987,Kerr1990,Luo2011,Luo2017,MacNeil1984,Scott1984,Scott1985,Scott1992,GeorgievGlazebrook2019,GeorgievGlazebrook2019b}.

To gain insights into the transport and utilization of energy by proteins,
in this work we have studied the quantum dynamics of multiple amide
I quanta, which arises from solving the Schr\"{o}dinger equation for the
generalized Davydov Hamiltonian \eqref{eq:Hamiltonian}. Utilizing
only standard quantum mechanical commutators and the Schrödinger equation,
we initially derived a discrete system of generalized Davydov equations
\eqref{eq:gauge-1} and \eqref{eq:gauge-2} that govern the quantum
dynamics of amide~I excitons in protein $\alpha$-helices (Section
\ref{sec:3}). Then, we performed computational simulations,
which have revealed that the discrete system of Davydov equations
supports the corresponding solitons, even for ultrashort protein $\alpha$-helices
whose length is only 10 peptide groups (Section \ref{sub:4-2}). This
suggests that natural evolution is able to select optimal protein
$\alpha$-helix lengths in gradual steps without losing the soliton
mechanism. Next, we have found that soliton collisions could lead
to either persistent pinned solitons, or to generating intermittent peaks of concentrated
amide~I energy through constructive quantum interference (Section
\ref{sub:4-3}). This provides a quantum physical mechanism for delivery
of concentrated peaks of energy at protein active centers.
The free energy delivered by the soliton to the active protein center may then be utilized to do physical work triggering chemical reactions and classical processes on nano-, micro- or milli-second timescales. The physical mechanisms for amplification of individual quantum processes to trigger macroscopic classical events at these slower timescales, which will effectively constitute the quantum-to-classical transition, were not covered in the current study and deserve further theoretical modeling along the lines of macromolecular electron clouds interacting with each other or with the electromagnetic field in biological systems \cite{Poznanski2019,Brizhik2010,Brizhik2015}.
Elaborate macro-quantum modeling of multiple proteins and their interaction with the quantized electromagnetic field would be particularly relevant for the function of protein voltage-gated ion channels incorporated in neuronal plasma membranes, where the electric field may reach values on the order of $10^7$ V/m during the course of action potentials fired by the neurons \cite{Georgiev2015,Georgiev2017}.

In addition to participation in quantum interference phenomena, the quantum nature of Davydov solitons was also illustrated by their capability to tunnel through massive barriers applied by external protein clamps
(Section \ref{sub:4-4}). This allows proteins to accomplish otherwise
classically impossible tasks. Lastly, we have demonstrated the importance
of the biological order for soliton stability through the regular
helical geometry that ensures uniform dipole--dipole coupling energies
between neighboring amide~I oscillators (Section \ref{sub:4-5}).
Interestingly, the presence of disorder in the dipole--dipole coupling
energies did not cause direct soliton disintegration, but instead resulted in compartmentalization
of the solitons within protein segments flanked with low dipole--dipole
couplings. This may have been valuable for the natural evolution of
protein function because random coil domains could evolve into $\alpha$-helices,
or vice versa, thereby sculpturing-out either energy transmission lines
or protein active centers for utilization of the delivered energy.

The present findings may also shed new light on the abiotic origin
of life where randomly assembled polypeptides could have found a way
to preserve existing seeds of order in the first replicators. Of course,
the origin of life is the most challenging problem of all time \cite{Pross2013}
and we may never find out how exactly life started as a consequence
of the very nature of the evolutionary process---we can only see the
best survivors having lost forever all those less fit ancestors that
were replaced in natural history. From living organisms that can fossilize,
we have been lucky to have a glimpse at what life was in past eras.
To look further back to the dawn of precellular life with the first
biomolecular replicators, however, is an uncertain task since single biomolecules
leave no traces, and the conditions on the early surface of the planet (Earth) are likely to have been
quite different from those which we currently imagine \cite{Kitadai2018}. There is evidence, however,
that a network of cross-replicating molecules is more robust, and operates
faster than a single self-replicator \cite{Lee1997,Duim2017}. Regardless
of whether polypeptides were present from the outset of life \cite{Andras2005}, or at a later stage joined up with
the biomolecular repertoire of living systems, our
results on the quantum dynamics of Davydov solitons in protein $\alpha$-helices
provide evidence for the non-trivial role of quantum effects in maintaining
life, and therefore highlight the prominent role of biological order to foster
those quantum effects. These events also contend that the first replicators
should have had lines for highly efficient quantum transmission of
energy, capable of trapping energy quanta in active centers, and the prowess
to utilize seeds of order to sustain non-trivial quantum dynamics
at biochemically relevant timescales.

\end{document}